\RequirePackage{ifpdf}
\pdfoutput=1
\documentclass[11pt,a4paper]{article}
\usepackage{jheppub}
\usepackage{graphicx}
\usepackage{amssymb}
\usepackage{amsmath}
\usepackage{amsthm}
\usepackage{slashed}
\usepackage{color,rotating}
\usepackage{bbm}
\usepackage{array}
\usepackage{colonequals}
\usepackage[utf8]{inputenc}
\DeclareMathOperator{\tr}{tr}
\allowdisplaybreaks

\newcommand{\UV}{{\small UV}}
\newcommand{\IR}{{\small IR}}

\newcommand{\RG}{{\small RG}}

\newcommand{\CHT}{{\small CHT}}
\newcommand{\CDT}{{\small CDT}}

\newcommand{\eg}{{\textit{e.g.}}}
\newcommand{\ie}{{\textit{i.e.}}}
\newcommand{\hmat}{\mathbbm h}
\newcommand{\hTL}{\mathbbm h^\text{TL}}
\newcommand{\hTr}{\mathbf h}

\graphicspath{{figures/}}

\title{Infinite order quantum-gravitational correlations}
\author{Benjamin Knorr}
\affiliation{Theoretisch-Physikalisches Institut, \\ Friedrich-Schiller-Universit{\"a}t
Jena,
Max-Wien-Platz 1, 07743 Jena, Germany}
\affiliation{Institute for Mathematics, Astrophysics and Particle Physics (IMAPP), \\
Radboud University Nijmegen, Heyendaalseweg 135, 6525 AJ Nijmegen, The Netherlands}
\emailAdd{b.knorr@science.ru.nl}

\abstract{A new approximation scheme for nonperturbative renormalisation group equations for quantum gravity is introduced.
Correlation functions of arbitrarily high order can be studied
by resolving the full dependence of the renormalisation group equations on the fluctuation field (graviton).
This is reminiscent of a local potential approximation in O(N)-symmetric field theories. As a first proof of principle, we derive
the flow equation for the ``graviton potential'' induced by a conformal fluctuation and corrections induced by a gravitational wave fluctuation.
Indications are found that quantum gravity might be in a non-metric phase in the deep ultraviolet.
The present setup significantly improves the quality of previous fluctuation vertex studies by including infinitely many couplings, thereby testing the reliability of schemes to 
identify different couplings to close the equations, and represents an important step towards the resolution of the Nielsen identity. The setup further 
allows in principle to address the question of putative gravitational condensates.
}

\keywords{Models of Quantum Gravity, Renormalisation Group, Nonperturbative Effects, Nonperturbative Renormalisation}

\begin{document}
%======================================================
\maketitle
%======================================================
\section{Introduction}
%======================================================

Einstein's metric theory of gravity is known not to be perturbatively renormalisable. Many different approaches
to overcome this difficulty were proposed in the past decades. One remarkably economic proposal is due to Weinberg \cite{Weinberg:1980gg},
he conjectured that gravity might be nonperturbatively renormalisable by an interacting fixed point of
its renormalisation group flow. If the critical hypersurface of this fixed point has a finite dimension, the resulting
theory is as predictive as an asymptotically free theory.

Only in the 90s, with the advent of functional renormalisation group (\RG{}) equations \cite{Wetterich:1992yh, Morris:1993qb, Reuter:1996cp}, this proposal 
could be tested in $d=4$.
Since then, a plethora of approximations and aspects were studied, including different approximations on the Einstein-Hilbert sector \cite{Falkenberg:1996bq, 
Souma:1999at, Lauscher:2001ya, Lauscher:2001rz, Reuter:2001ag, Litim:2003vp, Lauscher:2005qz, Reuter:2005bb, Niedermaier:2006wt, Groh:2010ta, Benedetti:2010nr, 
Manrique:2011jc, Reuter:2012id, Harst:2012ni, Litim:2012vz, Nink:2012kr, Rechenberger:2012dt, Nink:2014yya, Gies:2015tca, Falls:2015qga, Falls:2015cta}, higher 
derivatives and $f(R)$
\cite{Lauscher:2002sq, Codello:2006in, Benedetti:2009rx, Groh:2011vn, Rechenberger:2012pm, Ohta:2013uca, Machado:2007ea, Codello:2007bd, Bonanno:2010bt, 
Demmel:2012ub, Dietz:2012ic, Falls:2013bv, Dietz:2013sba, Falls:2014tra, Demmel:2014hla, Demmel:2014sga, Eichhorn:2015bna, Demmel:2015oqa, Ohta:2015efa, 
Ohta:2015zwa, Ohta:2015fcu, 
Falls:2016wsa, Falls:2016msz, Morris:2016spn, Christiansen:2016sjn, Gonzalez-Martin:2017gza, Hamada:2017rvn, Nagy:2017zvc}, the two-loop counterterm 
\cite{Gies:2016con}, 
aspects of unitarity \cite{Nink:2015lmq, Becker:2017tcx}, different variables 
\cite{Manrique:2011jc, Rechenberger:2012dt, Biemans:2016rvp, Biemans:2017zca, Houthoff:2017oam} and the coupling to matter \cite{Percacci:2002ie, 
Percacci:2003jz, Zanusso:2009bs, Daum:2009dn, Narain:2009fy, Manrique:2010mq, Vacca:2010mj, Harst:2011zx, Eichhorn:2011pc, Folkerts:2011jz, Dona:2012am, 
Dobrich:2012nv, Eichhorn:2012va, Dona:2013qba, Henz:2013oxa, Eichhorn:2014qka, Percacci:2015wwa, Borchardt:2015rxa, Dona:2015tnf, Labus:2015ska, 
Meibohm:2015twa, Eichhorn:2016esv, Meibohm:2016mkp, Eichhorn:2016vvy, Henz:2016aoh, Christiansen:2017gtg, Biemans:2017zca, Christiansen:2017qca, 
Eichhorn:2017eht, Eichhorn:2017ylw, Wetterich:2017ixo, Eichhorn:2017lry, Eichhorn:2017sok, Christiansen:2017cxa}.
All pure gravity studies so far are compatible with a fixed point that can control the ultraviolet (\UV{}) behaviour of quantum gravity, and most works
that also studied the inclusion of matter arrive at a similar conclusion for matter content compatible with the Standard Model.

A central technical tool to investigate gravitational (or gauge) \RG{} equations is the background field method. In the context of functional \RG{}
equations, severe problems arise from that, since the regularisation breaks the split Ward identity, as it depends on the background and the fluctuation
separately. It is known that if not treated with enough care, this can even change the universal 1-loop beta function
in Yang-Mills theory \cite{Litim:2002ce, Folkerts:2011jz}, or destroy the well-known Wilson-Fisher fixed point of the Ising model in $d=3$ 
\cite{Bridle:2013sra}. 
A modified version of the split Ward identity exists, which accommodates this deficit \cite{Reuter:1996cp, 
Pawlowski:2003sk, Pawlowski:2005xe, Manrique:2009uh, Manrique:2010mq, Donkin:2012ud, Bridle:2013sra, Dietz:2015owa, Safari:2015dva, Labus:2016lkh, 
Morris:2016nda, Safari:2016dwj, Safari:2016gtj, Wetterich:2016ewc, Morris:2016spn, Percacci:2016arh, Ohta:2017dsq, Nieto:2017ddk},
but is inherently difficult to solve. A more hands-on approach is to resolve the dependence of 
the effective action on both the background and the fluctuation field, which was systematically developed and applied to quantum gravity in 
\cite{Christiansen:2012rx, Codello:2013fpa, Christiansen:2014raa, Christiansen:2015rva, Dona:2015tnf, Meibohm:2015twa, Denz:2016qks, Meibohm:2016mkp, 
Henz:2016aoh, Christiansen:2016sjn, Knorr:2017fus, Christiansen:2017cxa}, for related approaches see also \cite{Manrique:2009uh, Manrique:2010mq, Manrique:2010am, Becker:2014qya, 
Becker:2014jua, Donkin:2012ud, Demmel:2014hla, Wetterich:2016ewc, Wetterich:2016qee}.

So far in this bimetric setting, it was only possible to derive beta functions for a finite number of couplings. To close these equations,
some couplings had to be identified. A priori, it is not clear how this should be done, and whether such a procedure is as stable as in
\eg{} scalar field theories, where one can find the Wilson-Fisher fixed point by a (low order) Taylor expansion of the potential around the vacuum expectation 
value with accurate estimates for the first critical exponent. The aim of this work is to develop the techniques to lift this restriction, 
and thus to treat an arbitrary dependence on the fluctuation field.
With this it is then possible to study whether the graviton acquires a nonvanishing vacuum expectation value, \ie{} if there is gravitational condensation.

This work is structured as follows: in Section \ref{sec:FlucDep} we introduce the basic ideas of our approach and some technical prerequisites necessary for 
the subsequent discussion. Section \ref{sec:LPA} contains the setup and approximations that we use to study the \UV{} structure of quantum gravity. In section 
\ref{sec:results} we present the numerical results for the fixed point structure, followed by a discussion of the physical significance of these results and 
potential shortcomings of the approximations in section \ref{sec:discussion}. We conclude with a short summary in section \ref{sec:summary}. The appendices 
collect some technical details.

%======================================================
\section{Functions of the fluctuation field}\label{sec:FlucDep}
%======================================================

Lagrangian formulations of the quantisation of gravity often rely on the background field method. For this,
the full metric $g$, which contains all information, is split into a nondynamical background metric $\bar g$,
and quantum fluctuations around this background, parameterised by a symmetric tensor field $h$. Many calculations
rely on a so-called linear split or parameterisation,
\begin{equation}
 g_{\mu\nu} = \bar g_{\mu\nu} + h_{\mu\nu} \, .
\end{equation}
More recently, other types of parameterisation have been studied \cite{Nink:2014yya, Demmel:2015zfa, Percacci:2015wwa,
Gies:2015tca, Labus:2015ska, Ohta:2015zwa, Ohta:2015efa, Falls:2015qga, Ohta:2015fcu, Dona:2015tnf, Ohta:2016npm, Falls:2016msz, Percacci:2016arh}, \eg{} the 
exponential split
\begin{equation}
 g_{\mu\nu} = \bar g_{\mu\rho} \, {\left( \exp (\bar g^{-1} h) \right)^\rho}_\nu \equiv \bar g_{\mu\rho} {\left( \exp (\hmat) \right)^\rho}_\nu \, ,
\end{equation}
which plays a distinguished role in two-dimensional quantum gravity \cite{Kawai:1993fq,Kawai:1992np,
Kawai:1993mb,Kawai:1995ju,Aida:1994zc}, and in general has the virtue of
being a one-to-one mapping between metrics $g$ and symmetric fluctuation tensors $h$ \cite{Demmel:2015zfa}.
Here, we introduced the $(1,1)$-tensor $\hmat = \bar g^{-1} h$ for convenience. The special role of $\hmat$ in the exponential parameterisation was 
already pointed out in \cite{Percacci:2016arh}, where it was called $X$. For us, $\hmat$ is useful since powers of it are automatically background covariant. It is also 
useful to introduce a traceless decomposition,
\begin{equation}
 \hmat = \hTL + \frac{1}{4} \hTr \, \mathbbm 1_4 \, ,
\end{equation}
where $\hTr$ is the trace of $\hmat$. Here, we already specified four spacetime dimensions, although the following discussion easily extends to other dimensions.

We will limit our discussion in this work to scalar invariants of $\hmat$ under $GL(4)$, \ie{} we don't consider invariants built with derivatives. Since 
$\hmat$ can be interpreted as a usual matrix, the scalar invariants are exactly the four eigenvalues. Clearly, this is of minor use in a functional language, 
and we have to find a useful way to form invariants. For this task, we use the Cayley-Hamilton theorem (\CHT{}), which is reviewed briefly in appendix 
\ref{sec:CHT}. Essentially, it states that, if we replace the eigenvalue by the matrix itself in the characteristic equation of the matrix, we get the zero 
matrix. For $\hTL$, it reads
\begin{equation}
 \left[\hTL\right]^4 - \frac{1}{2} \tr \left(\left[\hTL\right]^2\right) \left[\hTL\right]^2 - \frac{1}{3} \tr \left(\left[\hTL\right]^3\right) \hTL + \det \left( \hTL \right) \, \mathbbm{1}_4 = 0 \, .
\end{equation}
This immediately gives a basis of monomials - the four lowest nonnegative powers of the matrix - and an algorithm to expand any higher power of the 
matrix to a linear combination of these low powers. Scalar invariants of the matrix are then the determinant and the traces of the first three powers of the 
matrix. Employing the traceless decomposition, the invariants are
\begin{equation}\label{eq:invars}
\begin{aligned}
 \mathfrak{h}_1 &= \hTr \, , \\
 \mathfrak{h}_2 &= \tr \left( \left[ \hTL \right] ^2 \right) \, , \\
 \mathfrak{h}_3 &= \tr \left( \left[ \hTL \right]^3 \right) \, , \\
 \mathfrak{h}_4 &= \det \left( \hTL \right) \, .
\end{aligned}
\end{equation}
Thus, the most general parameterisation of the full metric which doesn't introduce a scale or uses derivatives can be written as
\begin{equation}\label{eq:metric_fulltext}
 g = \bar g \left( \mathcal A_0 \, \mathbbm 1_4  + \mathcal A_1 \, \hTL + \mathcal A_2 \left[ \hTL \right]^2 + \mathcal A_3 \left[ \hTL \right]^3 \right) \, .
\end{equation}
No higher powers of $\hTL$ appear since by the \CHT{} they can be reduced to lower powers, and are thus included in this ansatz.
The functions $\mathcal A_i$ are free functions of the four invariants \eqref{eq:invars}, and determine the parameterisation. The only constraints on them
are that $g$ should be invertible, \ie{} $\det g \neq 0$, and $g = \bar g$ if $h=0$, which fixes $\mathcal A_0 = 1 + \mathcal O(h)$. The inverse 
metric $g^{-1}$ has a similar exact representation,
\begin{equation}\label{eq:invmetric_fulltext}
 g^{-1} = \left( \mathcal B_0 \, \mathbbm 1_4  + \mathcal B_1 \, \hTL + \mathcal B_2 \left[ \hTL \right]^2 + \mathcal B_3 \left[ \mathbbm 
h^\text{TL} \right]^3 \right) \bar g^{-1} \, .
\end{equation}
The functions $\mathcal B_i$ can be expressed explicitly in terms of the $\mathcal A_i$, the full expressions are collected in appendix 
\ref{sec:paramandinverse}. As an example, for the linear split,
\begin{equation}\label{eq:linsplit_explicit}
 \mathcal A_0 = 1 + \frac{1}{4} \mathfrak{h}_1 \, , \quad \mathcal A_1 = 1 \, , \quad \mathcal A_2 = \mathcal A_3 = 0 \, ,
\end{equation}
and we find
\begin{equation}\label{eq:grav_inversecoeffs}
\begin{aligned}
 \mathcal B_0 &= \frac{\mathcal A_0^3 - \frac{1}{2} 
\mathcal A_0 \mathfrak{h}_2 + \frac{1}{3} \mathfrak{h}_3}{
\mathcal A_0^4 - \frac{1}{2} \mathcal A_0^2 
\mathfrak{h}_2 + \frac{1}{3} \mathcal A_0 \mathfrak{h}_3 + 
\mathfrak{h}_4} \, , \\
 \mathcal B_1 &= -\frac{\mathcal A_0^2 - \frac{1}{2} 
\mathfrak{h}_2}{\mathcal A_0^4 - \frac{1}{2} 
\mathcal A_0^2 \mathfrak{h}_2 + \frac{1}{3} \mathcal 
A_0 \mathfrak{h}_3 + \mathfrak{h}_4} \, , \\
 \mathcal B_2 &= \frac{\mathcal A_0}{\mathcal 
A_0^4 - \frac{1}{2} \mathcal A_0^2 \mathfrak{h}_2 + 
\frac{1}{3} \mathcal A_0 \mathfrak{h}_3 + \mathfrak{h}_4} \, , \\
 \mathcal B_3 &= -\frac{1}{\mathcal A_0^4 - \frac{1}{2} 
\mathcal A_0^2 \mathfrak{h}_2 + \frac{1}{3} \mathcal 
A_0 \mathfrak{h}_3 + \mathfrak{h}_4} \, .
\end{aligned}
\end{equation}
It is further straightforward to calculate the determinant of the general metric \eqref{eq:metric_fulltext}. Again, the full expression is deferred to the 
appendix. For the linear split \eqref{eq:linsplit_explicit},
\begin{equation}\label{eq:detglinsplit}
 \det g = \left[ \mathcal A_0^4 - \frac{1}{2} \mathcal A_0^2 \mathfrak{h}_2 + \frac{1}{3} \mathcal A_0 \mathfrak{h}_3 + \mathfrak{h}_4 \right] \det \bar g \, .
\end{equation}
Let us stress again that the expressions just presented are \textit{exact}, and follow directly from the \CHT{}. In appendix \ref{sec:paramandinverse} we 
collect some formulas for the exponential split.

It is in fact advantageous to use the traceless decomposition to define the scalars $\mathfrak{h}_i$, because they are
in some sense orthogonal, which makes it possible to choose fluctuations $h$ such that only some of the invariants have a nonvanishing value.
This comes in useful if one wants to employ approximations, where one only considers the dependence on a subset of these
scalars. As an example, a gravitational wave fluctuation,
\begin{equation}\label{eq:GW}
 \hTL = 
 \left(
 \begin{array}{cccc}
  0 & 0 & 0 & 0 \\
  0 & h_+ & h_\times & 0 \\
  0 & h_\times & -h_+ & 0 \\
  0 & 0 & 0 & 0
 \end{array}
 \right) \, ,
\end{equation}
gives
\begin{equation}
\begin{aligned}
 \mathfrak{h}_2 &= 2 \left( h_+^2 + h_\times^2 \right) \, , \\
 \mathfrak{h}_3 &= 0 \, , \\
 \mathfrak{h}_4 &= 0 \, .
\end{aligned}
\end{equation}
Here, $h_+$ and $h_\times$ are the two polarisation states of the gravitational wave in transverse traceless gauge. As claimed, it is thus easily possible
to truncate the invariants \eqref{eq:invars} by restricting the considered fluctuations to special choices.
One should also notice that $\mathfrak{h}_2 \geq 0$, and moreover $\mathfrak{h}_2=0$ implies $\mathfrak{h}_3=\mathfrak{h}_4=0$,
which can easily be seen if the scalars are expressed in terms of the eigenvalues of the matrix $\hTL$.

%======================================================
\section{Functional RG and local potential approximation in quantum gravity}\label{sec:LPA}
%======================================================

In this work we use the functional renormalisation group to calculate nonperturbative
beta functions. The central object of study is the effective average action which interpolates
between the classical action in the ultraviolet (\UV{}) and the standard effective action in the infrared (\IR{}),
modulo some subtleties \cite{Morris:1993qb, Manrique:2008zw, Manrique:2009tj, Morris:2015oca}. It is regularised
by a momentum-dependent effective mass term, which renders any infinitesimal \RG{} step finite.
This follows the Wilsonian idea of integrating out momentum modes shell by shell.
The effective average action (or effective action for short in the following) in our setup is
a functional of both the background metric and the fluctuation field, $\Gamma = \Gamma[\bar g; h]$.
The individual dependence on $\bar g$ and $h$ is necessary to define the regulator
and the gauge fixing, and thus gives rise to the split-Ward identity. Still, background diffeomorphism
invariance can be explicitly maintained. The \RG{} flow of the effective action is governed by
the exact flow equation \cite{Wetterich:1992yh, Morris:1993qb, Reuter:1996cp}
\begin{equation}\label{eq:floweq}
 k \partial_k \Gamma \equiv \dot{\Gamma} = \frac{1}{2} \text{STr} \left[ \left( \Gamma^{(2)} + \mathfrak R \right)^{-1} \dot{\mathfrak{R}} \right] 
\, .
\end{equation}
Here, $k$ is the \IR{} cutoff scale, an overdot indicates $k$ times the derivative w.r.t.\ $k$, $\mathfrak R$ is the regulator kernel, STr stands for the 
supertrace, summing over all dynamical fields and discrete indices, integrating over continuous indices and
multiplying a minus sign for Gra\ss{}mann-valued fields, and $\Gamma^{(2)}$ is the Hessian
of the effective action w.r.t.\ the fluctuation field(s). For reviews of the flow equation in quantum gravity see
\eg{} \cite{Reuter:1996ub,Pawlowski:2005xe,Niedermaier:2006wt,Percacci:2007sz,Reuter:2012id,Nagy:2012ef}.

\subsection{Nielsen identity}

The standard effective action, given by the effective average action in the limit $k\to0$, can only depend on one field, the metric, \ie{} it is diffeomorphism invariant.
For finite $k$, this is broken by the regulator and the gauge-fixing.
The amount of breaking can be expressed by the so-called Nielsen or split-Ward identity.
It relates the derivatives of the effective action w.r.t.\ the fluctuation field to the ones
w.r.t.\ the background metric. Schematically, it reads
\begin{equation}
 \frac{\delta \Gamma}{\delta \bar g} - \frac{\delta \Gamma}{\delta h} = \mathcal R + \mathcal G \, ,
\end{equation}
where $\mathcal R$ and $\mathcal G$ are the breaking terms arising from the regulator and the gauge fixing, respectively, which depend on both the background metric
and the fluctuation field individually, more specifically \textit{not} in the combination of the full metric.
An introduction to the most important
points can be found in \cite{Denz:2016qks}, for more general discussions, see \eg{} \cite{Reuter:1996cp, Pawlowski:2003sk,
Pawlowski:2005xe, Manrique:2009uh, Manrique:2010mq, Donkin:2012ud, Bridle:2013sra, Dietz:2015owa, Safari:2015dva, Morris:2016spn,
Percacci:2016arh, Ohta:2017dsq, Nieto:2017ddk}. For us, the central
observation is that the background independence of observables and the nontrivial Ward identity
necessitate the direct computation of fluctuation correlation functions.

So far, this direct computation was done in a vertex expansion, both for pure gravity \cite{Christiansen:2012rx,
Codello:2013fpa, Christiansen:2014raa, Christiansen:2015rva, Denz:2016qks, Christiansen:2016sjn, Knorr:2017fus}
as well as gravity-matter systems \cite{Meibohm:2015twa, Dona:2015tnf, Meibohm:2016mkp, Eichhorn:2016esv, Eichhorn:2017sok, Christiansen:2017cxa}.
Related approaches are bimetric calculations \cite{Manrique:2009uh, Manrique:2010mq, Manrique:2010am, Becker:2014qya, Becker:2014jua} and
efforts to solve the Nielsen identity
directly or indirectly \cite{Pawlowski:2003sk, Donkin:2012ud, Dietz:2015owa, Safari:2015dva, Labus:2016lkh,
Morris:2016nda, Safari:2016dwj, Safari:2016gtj, Wetterich:2016ewc, Morris:2016spn, Percacci:2016arh, Ohta:2017dsq, Nieto:2017ddk}.
The most advanced vertex calculation on a flat background resolved parts of the 
four-point function \cite{Denz:2016qks}, whereas in \cite{Knorr:2017fus}, fluctuation-curvature-correlations were resolved for the first time. All of these 
calculations rely on coupling identifications of higher
vertices to close the flow equation \eqref{eq:floweq}. With the scheme that we present below, we can overcome this deficit.
For the first time, this gives direct access to correlation functions of arbitrary order, and a means to check in how much the coupling identifications are a stable approximation. By this we also 
automatically provide the explicit mapping between the ``bimetric'' and the ``fluctuation'' language, \ie{} the different possibilities between spanning the 
effective action with the background and the full metric, or with the background metric and the fluctuation field.
The remainder of this section is devoted to introduce our approximation in which we solve the flow equation \eqref{eq:floweq}.

\subsection{Einstein-Hilbert part}\label{subsec:EH}

Our ansatz for the kinetic and potential part of the effective action reads
\begin{equation}\label{eq:EHansatz}
 \Gamma_\text{fluc} = \frac{1}{16\pi G_N} \int \text{d}^4x \, \left( - \sqrt{\det g} \, R + \sqrt{\det \bar g} \, 2\mathcal V(\mathfrak{h}_1, \mathfrak{h}_2, 
\mathfrak{h}_3, 
\mathfrak{h}_4) \right) \, .
\end{equation}
In this ansatz, $G_N$ is the (running) Newton's constant, $R$ is the Ricci scalar of the full metric $g$ and $\mathcal V$ is the fluctuation potential,
which is also $k$ dependent. In previous studies, the latter part was approximated by the classical Einstein-Hilbert
structure, partly resolving different vertex couplings (conventionally called $\lambda_i$ for the coupling
of the $i$-th vertex). For the first time, we go beyond this,
resolving the full fluctuation field dependence of the constant part of all vertices.
This also extends earlier similar work in conformally reduced gravity \cite{Reuter:2008wj, Reuter:2008qx, Machado:2009ph, Bonanno:2012dg, Dietz:2015owa, 
Labus:2016lkh, Dietz:2016gzg}.
The goal of this work is to derive and solve the beta function for $\mathcal V$ in some approximation.

\subsection{Gauge-fixing and ghosts}\label{subsec:GFandgh}

The action \eqref{eq:EHansatz} has to be complemented by a gauge fixing term,
\begin{equation}
 \Gamma_\text{gf} = \frac{1}{32\pi G_N \alpha} \int \text{d}^4 x \, \sqrt{\det \bar g} \, \bar g^{\mu\nu} F_\mu F_\nu \, ,
\end{equation}
with the gauge fixing condition
\begin{equation}
 F_\mu = \mathcal F_\mu^{\rho\sigma}[\bar g] h_{\rho\sigma} = \left( \delta^\beta_\mu \bar D^\alpha
 - \frac{1+\beta}{4} \bar g^{\rho\sigma} \bar D_\mu \right) h_{\rho\sigma} \, .
\end{equation}
By $\bar D$ we understand the covariant derivative constructed from the background metric $\bar g$, and $\alpha$ and $\beta$ are gauge fixing parameters.
Eventually, we are interested in the Landau limit $\alpha\to0$ which implements the gauge fixing strictly. In Landau gauge, neither of the gauge fixing 
parameters flows, no matter the value of $\beta$ \cite{Knorr:2017fus, Litim:1998qi}.

With the introduction of a gauge fixing, we have to account for the corresponding change in the measure by introducing
ghost fields. The resulting ghost action reads
\begin{equation}
 \Gamma_\text{gh} = \int \text{d}^4 x \, \sqrt{\det \bar g} \, \bar c_\mu \bar g^{\mu\nu} \mathcal F_\nu^{\rho\sigma}[\bar g] \delta_Q h_{\rho\sigma} \, .
\end{equation}
Since we gauge-fix $h$ directly, independently of the parameterisation, the quantum gauge transformation $\delta_Q h$ is in general nontrivial, and not 
simply given by the Lie derivative of the full metric along the ghost vector field $c$. Only for the special case of a linear parameterisation, we arrive
at the familiar form
\begin{equation}
 \delta_Q h_{\mu\nu} = \delta_Q g_{\mu\nu} = \mathcal L_c g_{\mu\nu} = D_\mu c_\nu + D_\nu c_\mu \, .
\end{equation}
We show how to derive the relevant general expression in appendix \ref{sec:gf}. In particular, the quantum gauge transformation for the exponential
parameterisation is given in \eqref{eq:qu_ga_tr_exp}. In our investigations it turned out that if we gauge-fix the full metric
instead, the results rely heavily on the choice of the regulator.

\subsection{Regulators}\label{subsec:reg}

We finally have to fix the regularisation of both the gravitons and the ghost fields. In this, we closely follow the strategy of \cite{Christiansen:2012rx, 
Christiansen:2014raa, Christiansen:2015rva, Denz:2016qks, Christiansen:2016sjn, Knorr:2017fus}, choosing the regulator proportional to the two-point function, 
with the potential $\mathcal V$ and the background curvature $\bar R$ set to zero:
\begin{align}
 \Delta S_\text{grav} &= \frac{1}{2} \int \text{d}^4 x \, \sqrt{\det \bar g} \, h_{\mu\nu} \frac{\mathfrak R(\bar \Delta)}{\bar \Delta}
 \left[\left. \left( \Gamma_\text{fluc} + \Gamma_\text{gf} \right)^{(2)\mu\nu\rho\sigma} \right|_{\mathcal V=h=\bar R=0}\right] h_{\rho\sigma} \, , \\
 \Delta S_\text{gh} &= \int \text{d}^4 x \, \sqrt{\det \bar g} \, \bar c_\mu \frac{\mathfrak R(\bar \Delta)}{\bar \Delta} \bar g^{\mu\nu} \mathcal 
F_\nu^{\rho\sigma}[\bar g] \left. \mathcal \delta_Q h_{\rho\sigma} \right|_{h=\bar R=0} \, .
\end{align}
For convenience, we introduced the Laplacean of the background covariant derivative by $\bar \Delta = -\bar D^2$.
Since we gauge-fix $h$ instead of $g$, we can also employ the minimal regulator recently introduced in 
\cite{Knorr:2017fus}, which only changes the background part of the flow, \ie{} the part at vanishing fluctuation field, resulting in an overall shift of the 
potential $\mathcal V$. We checked explicitly that this is indeed the case to all orders in the fluctuation field in the truncation that we discuss subsequently.

\subsection{Flow equations}\label{subsec:flow}

Now, we are in the situation to calculate the flow equation for the fluctuation potential $\mathcal V$.
It is enough to use a flat background metric $\bar g = \eta$ after the hessian has been calculated. To simplify matters, we don't derive a flow equation for 
the Newton's constant, rather treating it as a parameter.
The huge amount of tensor algebra is dealt with by the Mathematica package \textit{xAct} 
\cite{xActwebpage, Brizuela:2008ra, 2008CoPhC.179..597M, 2007CoPhC.177..640M, 2008CoPhC.179..586M, 2014CoPhC.185.1719N}.

Before we carry on, let us discuss our choice of parameterisation. First consider the linear split. It turns out that the constraints on the positivity of the 
determinant of the metric severely impacts the accessible fluctuation space, and always gives rise to a singular line where $\det g=0$, see \eqref{eq:detglinsplit}, which 
cuts the space spanned by \eqref{eq:invars}. Put differently, there are fluctuations $h$ such that the full metric is degenerate or even has the opposite sign,
and these fluctuations should tentatively be excluded from the path integral.
Since it is a huge technical and numerical hurdle to treat the restricted variable space and the singular line, we will instead consider the 
exponential split in the following, since it doesn't give rise to any singularities for finite fluctuations due to being a one-to-one map between metrics and fluctuations.
We will further assume that the path integral measure is trivial for the exponential split, for a discussion of this see \cite{Gies:2015tca}.

It is clearly a formidable task to derive the beta function of the full fluctuation potential. As a proof of concept, we shall make a further approximation 
where we can derive the flow equation with manageable effort, and give comments on the quality and the potential impact of improvements of the approximation later. We thus 
restrict ourselves in the following to a trace fluctuation together with the leading order of a gravitational wave fluctuation. Our ansatz reads
\begin{equation}
 \Gamma_\text{trunc} = \frac{1}{16\pi G_N} \int \text{d}^4x \, \left[ - \sqrt{\det g} R + \sqrt{\det \bar g} \, 2 \left( \mathcal V(\mathfrak{h}_1) + 
\mathfrak h_2 \mathcal W(\mathfrak h_1) \right) \right] + \Gamma_\text{gf} + \Gamma_\text{gh} \, .
\end{equation}
To obtain the flow equations for $\mathcal V$ and $\mathcal W$, we take the full second functional derivative of this action, and only afterwards project onto
fluctuations which include the full $\mathfrak h_1$-dependence and all terms up to linear order in $\mathfrak h_2$.

A further technical problem arises when calculating the propagator and eventually the trace, which we mention before we finally present the explicit flow 
equations. The inversion of the regularised two-point function gives rise to terms where $\hTL$ is contracted with the momentum $p$, 
\eg{} $p^\mu h^\text{TL}_{\mu\nu} p^\nu$. These then appear in denominators, and we have to clarify how to integrate over the momenta,
eventually expressing them in terms that only involve the scalar invariants \eqref{eq:invars}. In the 
special case of a gravitational wave fluctuation, we can simply insert the explicit matrix representation \eqref{eq:GW}, which with our truncation can be 
rewritten as
\begin{equation}
 \hTL = \text{diag}\left( 0, \sqrt{\mathfrak h_2/2}, -\sqrt{\mathfrak h_2/2}, 0 \right) \, ,
\end{equation}
where we can without loss of generality set $h_\times=0$. With this, coordinates for the loop momentum can be chosen in order to calculate the loop integral. 
Still, in a more general setting, it is useful to have general formulas to treat these kinds of expressions. We will collect some aspects of this in appendix 
\ref{sec:loopint}.

To present the explicit flow equations, we first switch to dimensionless quantities by appropriate rescalings with powers of $k$,
\begin{equation}
 V = k^{-2} \mathcal V \, , \qquad W = k^{-2} \mathcal W \, , \qquad g = k^{2} G_N \, .
\end{equation}
The explicit flow equation for $V$ in the Landau limit $\alpha\to0$ then reads
\begin{align}\label{eq:flowV}
 \dot V(\mathfrak h_1) &= -4V(\mathfrak h_1) -\frac{g}{6 \pi  (\beta -3)^4 \left(1-e^{\frac{\mathfrak{h}_1}{4}}\right)^3} \Bigg(4 (\beta -3)^4 
\left(e^{\frac{\mathfrak{h}_1}{4}}-1\right)^3 +(\beta -3)^2 
\left(1-e^{\frac{\mathfrak{h}_1}{4}}\right) \times \notag \\
&\quad\left((\beta -3)^2 \left(-35 e^{\frac{\mathfrak{h}_1}{4}}+4
   e^{\frac{\mathfrak{h}_1}{2}}+13\right)+288 V''\left(\mathfrak{h}_1\right)-72 \left(\beta ^2-10 \beta +15\right) W\left(\mathfrak{h}_1\right)\right) \notag \\
   &\quad+3 
\left((\beta -3)^2-96 V''\left(\mathfrak{h}_1\right)-16 \beta ^2
   W\left(\mathfrak{h}_1\right)\right) \Big[ (\beta -3)^2 \left(2 e^{\frac{\mathfrak{h}_1}{4}}-1\right)-96 V''\left(\mathfrak{h}_1\right) \notag \\
   &\quad-16 \beta ^2 
W\left(\mathfrak{h}_1\right)\Big] \ln \left(\frac{(\beta -3)^2-96
   V''\left(\mathfrak{h}_1\right)-16 \beta ^2 W\left(\mathfrak{h}_1\right)}{(\beta -3)^2 e^{\frac{\mathfrak{h}_1}{4}}-96 V''\left(\mathfrak{h}_1\right)-16 
\beta 
^2 W\left(\mathfrak{h}_1\right)}\right) \notag \\
&\quad+15 (\beta -3)^4
   \left(8 W\left(\mathfrak{h}_1\right)+1\right) \left(2 e^{\frac{\mathfrak{h}_1}{4}}+8 W\left(\mathfrak{h}_1\right)-1\right) \ln \left(\frac{8 
W\left(\mathfrak{h}_1\right)+1}{e^{\frac{\mathfrak{h}_1}{4}}+8
   W\left(\mathfrak{h}_1\right)}\right)\Bigg) \, .
\end{align}
The flow equation for $W$ is even lengthier and will not be presented here.
In this, we assumed that the dimensionless Newton's constant 
$g$ is at an interacting fixed point, $\dot g =0$ and $g \neq 0$. We also chose the Litim regulator \cite{Litim:2001up, Litim:2002cf} to evaluate the integrals. Finally we shifted $V$ 
such that the quantum contribution to $\dot V$ vanishes in the limit $\mathfrak h_1\to\infty$ if $V=W=0$.
\begin{figure*}[t]
\begin{center}
\includegraphics[width=0.717\textwidth]{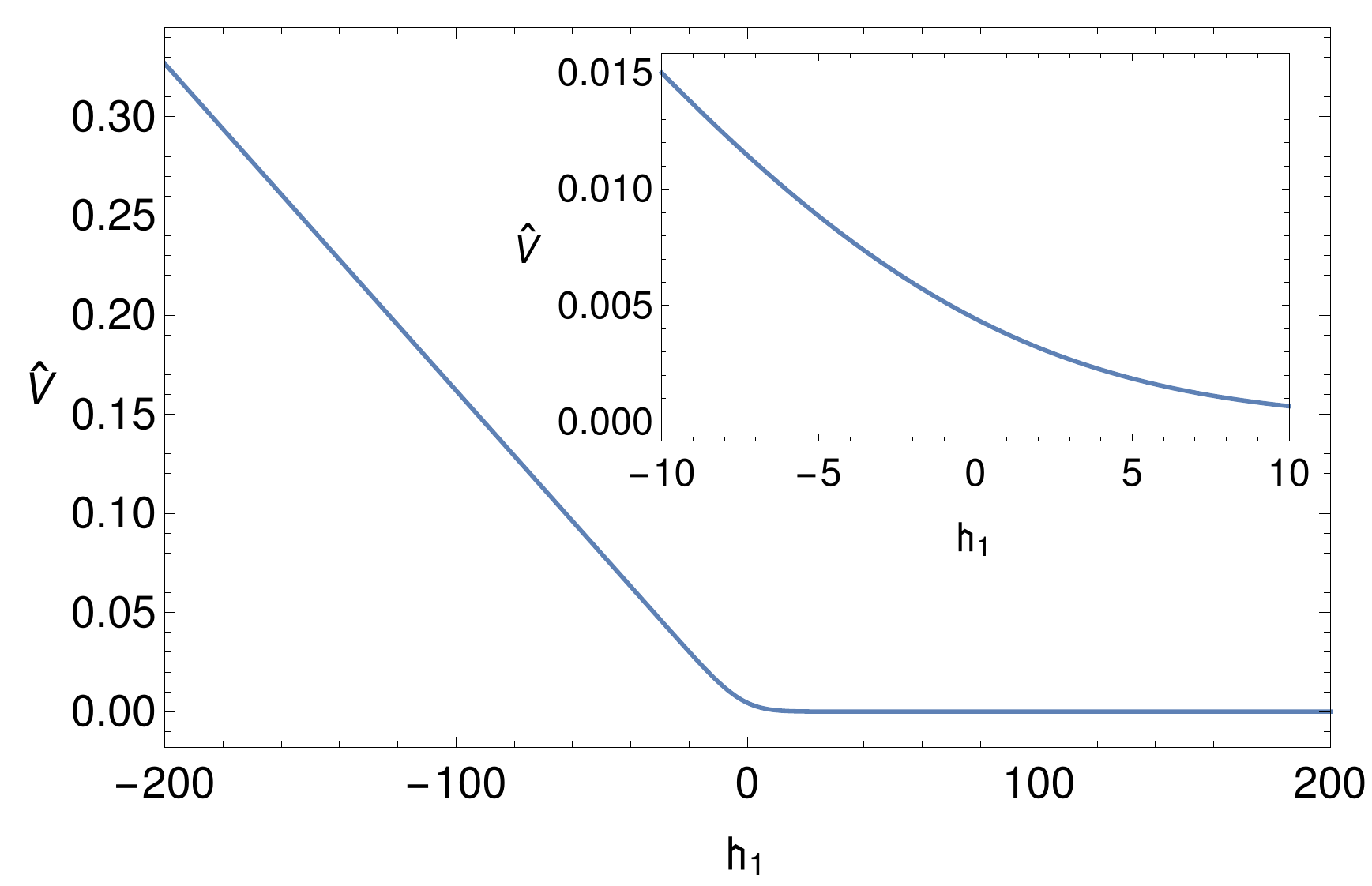}
\caption{Fixed point solution to \eqref{eq:flowVgaugeindependent} for the rescaled coupling $\hat g=1/36$, obtained with pseudo-spectral methods. The solution is a 
monotonically decreasing function, which rises linearly for large negative arguments, and drops exponentially for large positive arguments.}%
\label{fig:PSMsol1}
\end{center}
\end{figure*}
%

%======================================================
\section{Fixed point structure}\label{sec:results}
%======================================================

We will now study the fixed point structure of equation \eqref{eq:flowV} and the corresponding equation for $\dot W$. First, we discuss a 
truncation with $V$ alone, setting $W=0$, afterwards studying the coupled system. The subsequent numerical results are obtained with 
pseudo-spectral methods, which have been systematically developed in the context of functional \RG{} flows in \cite{Borchardt:2015rxa, Borchardt:2016pif}, and 
successfully been employed in \eg{} \cite{Heilmann:2014iga, Borchardt:2016xju, Borchardt:2016kco, Knorr:2016sfs, Knorr:2017yze}. As numerical parameters, we choose $g=1/4$ 
and $\beta=0$. The precise choice for these parameters is for purely illustrational purpose, but motivated by values obtained in recent studies \cite{Knorr:2017fus}.

\subsection{Conformal fluctuation potential}\label{subsec:Vonly}

If we set $W=0$ in \eqref{eq:flowV}, and further employ the rescalings $g\to(\beta-3)^2 \hat g$, $V\to(\beta-3)^2 \hat V$, all occurrences of the gauge 
parameter $\beta$ drop out. The flow equation for $\hat V$ is thus gauge independent,
\begin{equation}\label{eq:flowVgaugeindependent}
\begin{aligned}
 \dot{\hat V}(\mathfrak h_1) &= -4 \hat V(\mathfrak h_1) + \frac{\hat g}{2\pi \left( e^{\mathfrak h_1/4} - 1 \right)^3} \Bigg[ 3 + 96 \hat V''(\mathfrak h_1) + 9 e^{\mathfrak 
h_1/2} + \frac{5}{4}\mathfrak h_1 - \frac{1}{2} e^{\mathfrak h_1/4}(24+5\mathfrak h_1) \\
&\mkern-17mu -96e^{\mathfrak h_1/4}\hat V''(\mathfrak h_1) - \left( -1+2e^{\mathfrak h_1/4} - 96 \hat V''(\mathfrak h_1) \right) \left( -1+96 \hat V''(\mathfrak h_1) \right) 
\ln \frac{1-96 \hat V''(\mathfrak h_1)}{e^{\mathfrak h_1/4}-96 \hat V''(\mathfrak h_1)} \Bigg] \, .
\end{aligned}
\end{equation}
We find a single global solution for the fixed point condition $\dot{\hat V} = 0$, which is shown in Figure \ref{fig:PSMsol1}.
This solution is a monotonically decreasing function.
An asymptotic expansion around $\mathfrak h_1=-\infty$ is possible, where subleading terms are suppressed by powers of $(e^{\mathfrak h_1/4})$. The leading 
order is linear in $\mathfrak h_1$, in contrast to the naive expectation $\hat V\propto\sqrt{\det g/\det \bar g}=e^{\mathfrak h_1/2}$, thus we have strong 
fluctuation effects. Clearly, due to the $\beta$-dependent rescaling, all gauge dependence is hidden in the effective coupling $\hat g$. From recent studies 
\cite{Knorr:2017fus} we infer that this dependence is rather weak, signalling the stability of this result upon variations of $\beta$. The qualitative picture 
of the solution is already manifest if one expands $\hat V$ in powers of $\hat g$, keeping only few terms, and thus the solution varies essentially linearly
with $\hat g$, for small $\hat g$.

\subsection{Corrections by gravitational wave fluctuations}\label{subsec:gravwave}

\begin{figure*}[t]
\begin{center}
\includegraphics[width=\textwidth]{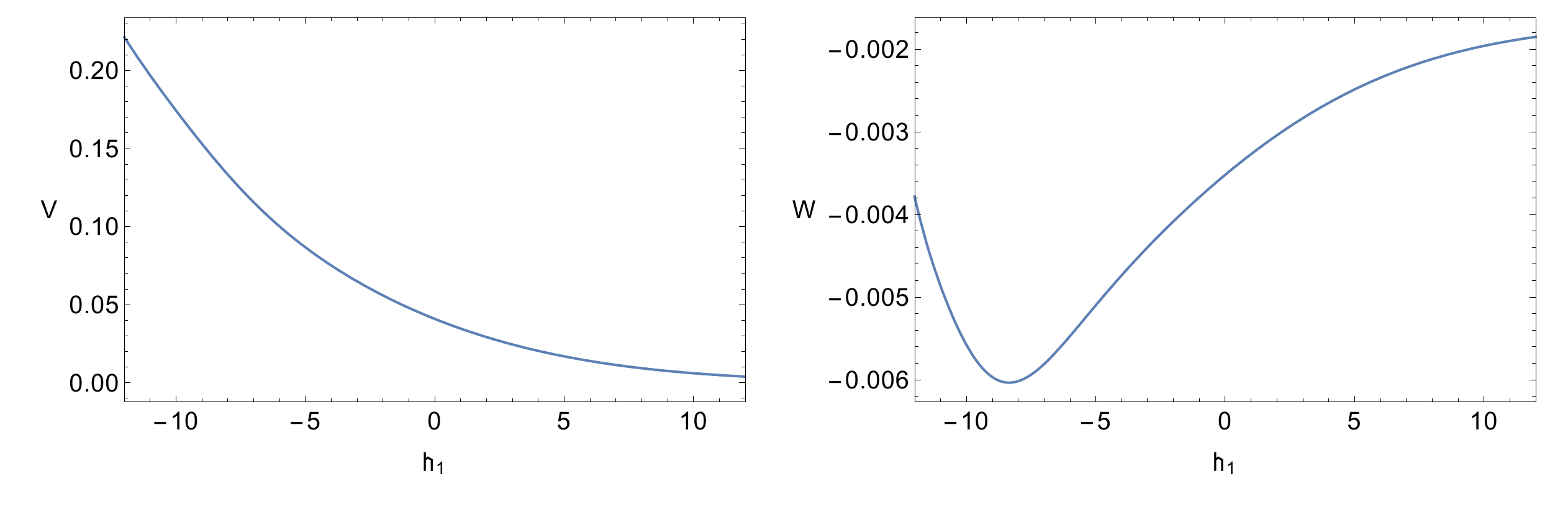}
\caption{Fixed point solution to the coupled system of flow equations for $V$ and $W$ for the coupling $g=1/4$ and the gauge fixing parameter $\beta=0$, obtained with 
pseudo-spectral methods. The qualitative picture for $V$ is the same as with $W=0$, which is reflected by the fact that $W$ itself is numerically small and 
only slowly varying.}%
\label{fig:PSMsol2}
\end{center}
\end{figure*}

We now discuss the coupled system of $V$ and $W$. Again, we find a single solution for the fixed point condition $\dot V = \dot W = 0$,
which is shown in Figure \ref{fig:PSMsol2}. The qualitative picture 
stays the same as discussed above: $V$ is monotonically decreasing, with similar asymptotics as with $W=0$. The difference in the absolute value is due to the fact that we 
didn't rescale $V$ as above, since the gauge dependence doesn't drop out in this extended approximation. On the other hand, the function $W$ is numerically 
small, and decreases exponentially in both limits $\mathfrak h_1 \to \pm\infty$. This property is not visible in Figure \ref{fig:PSMsol2} for large negative $\mathfrak h_1$.
The reason for this is that the coupled set of equations is numerically very difficult to solve for these values of $\mathfrak h_1$.

%======================================================
\section{Discussion}\label{sec:discussion}
%======================================================

We now interpret the results presented above and try to assess the reliability of the approximation. The first observation is that, since $V$ is 
monotonically decreasing with $V'(\infty)=0$, the natural expansion point around the minimum of the potential would be $\mathfrak h_1 = \infty$.
This immediately raises doubts whether the standard approach, namely an expansion around vanishing fluctuation field, is physically justified.
A potential explanation for this is that quantum gravity in the deep \UV{} is in a non-metric phase, and potentially more fundamental building blocks
as used in causal dynamical triangulations (\CDT{}) \cite{Ambjorn:1998xu, Ambjorn:2005db, Ambjorn:2006jf, Benedetti:2009ge, Anderson:2011bj,
Ambjorn:2011cg, Laiho:2011ya, Ambjorn:2012jv, Jordan:2013awa, Ambjorn:2013tki, Ambjorn:2013apa, Ambjorn:2014gsa, Cooperman:2014uha,
Ambjorn:2016mnn, Glaser:2016smx, Laiho:2016nlp, Ambjorn:2017tnl, Ambjorn:2017ogo, Glaser:2017yxz}
or causal sets \cite{Bombelli:1987aa, Sorkin:1990bj, Sorkin:2003bx, Surya:2011du, Surya:2011yh, Glaser:2013pca, Glaser:2014dwa, Eichhorn:2017djq, Glaser:2017sbe, Eichhorn:2017bwe}
are the true degrees of freedom.
On the other hand, we find that the linear correction due to gravitational wave fluctuations is strictly negative.
This could potentially ``cure'' the first observation, in the sense that the true minimum of the potential might be at a point with finite coordinates
$\mathfrak h_1, \mathfrak h_2 \neq 0$.
Second, the solution deviates strongly from the naive expectation involving the 
exponential parameterisation, \ie{} $V\propto\sqrt{\det g/\det \bar g}=e^{\mathfrak h_1/2}$. In particular, our solution rises linearly for large negative 
arguments, whereas it decreases exponentially fast for large positive arguments. This indicates very strong quantum fluctuations and emphasises the need for 
the present approach to resolve the full potential.

Let us now assess the quality of the present approximation. Since the graviton fluctuation is dimensionless, anomalous dimensions can play a very important 
role in the discussion of the fixed point structure. We will now try to analyse some scenarios that are possible if anomalous dimensions are included. For the 
discussion we will assume that $\mathfrak h_3$ and $\mathfrak h_4$ only play a subdominant role, thus we assume that $V(\mathfrak h_1,\mathfrak h_2)$ is a 
decent approximation to the true potential. The main impact of the anomalous dimensions $\eta_\text{Tr}$ and $\eta_\text{TL}$ of the trace and traceless mode, 
respectively, is the additional contribution to the canonical scaling,
\begin{equation}
 \dot V(\mathfrak h_1, \mathfrak h_2) = -4 V(\mathfrak h_1, \mathfrak h_2) + \frac{\eta_\text{Tr}}{2} \mathfrak h_1 \partial_{\mathfrak h_1} V(\mathfrak h_1, \mathfrak h_2) + 
\eta_\text{TL} \mathfrak h_2 \partial_{\mathfrak h_2} V(\mathfrak h_1, \mathfrak h_2) + \mathcal O(g) \, .
\end{equation}
Note that $\eta_\text{TL}$ is essentially the anomalous dimension of the physical transverse traceless mode, and its momentum dependence was calculated in 
\cite{Christiansen:2014raa}, with strictly positive sign and a value of $\mathcal O(1)$, although in a linear parameterisation
. For large $\mathfrak h_2$, the canonical scaling term indicates that $V\propto \mathfrak h_2^{4/\eta_\text{TL}}$ for $\mathfrak h_2\to\infty$ at the fixed point where $\dot V=0$.
On the other hand, for a well-defined propagator, we need that $\partial_{\mathfrak h_2} V>c_1$ with a finite constant 
$c_1$. This is the analogue of the singularity at $\Lambda=1/2$ in standard calculations. We thus conclude that for large $\mathfrak h_2$, the potential should 
rise like a power law. Together with the indications of the above results that $W<0$, this indeed strengthens the hint towards a minimum of the potential at
a finite value of $\mathfrak h_2$. Let us however stress that in this analysis we assumed that the quantum contribution is subleading in the limit of 
large $\mathfrak h_2$, which might not be the case.

Now we discuss the impact of $\eta_\text{Tr}$. Again assuming that the quantum term is subleading, we are lead to the conclusion that $V\propto \mathfrak 
h_1^{8/\eta_\text{Tr}}$ for large $|\mathfrak h_1|$ at the fixed point. On the other hand, this time we have an upper bound on the second derivative for a well-defined 
propagator, $\partial_{\mathfrak h_1}^2 V<c_2$. There are hence several distinct possibilities. If $\eta_\text{Tr}<0$, we conclude from the above results that for large negative 
$\mathfrak h_1$ actually the quantum term dominates, whereas for large positive arguments, the solution decreases by a power law instead of exponentially. 
Thus, the situation is qualitatively similar to the case analysed above. By contrast, if $0<\eta_\text{Tr}<2$, the asymptotic scaling is a power law with 
exponent larger than two, thus the prefactor is necessarily negative, and any putative fixed point potential is unbounded from below. It is not clear to the 
author how to interpret this case. One might argue that the trace mode is anyway not a propagating physical degree of freedom, and thus the physical part of 
the graviton potential isn't influenced by this unboundedness. Finally, if $\eta_\text{Tr}>2$, there is the possibility for a fixed point potential which is 
bounded from below and raising like a power law asymptotically. Necessarily, this gives rise to a minimum at a finite value of 
$\mathfrak h_1$. Together with the above observations that $W<0$ and $\eta_\text{TL}>0$, this gives a good chance that the potential admits a minimum at finite 
values of both invariants. Note however that such a large anomalous dimension could invalidate the standard way of regularisation, for 
a discussion of this aspect in quantum gravity coupled to matter, see \cite{Meibohm:2015twa}. The authors of \cite{Eichhorn:2017sok} reported small, but positive values for
$\eta_\text{Tr}$, however the calculation was done in a linear parameterisation, further neglecting the gaps of the graviton fluctuation. We thus cannot give a definite conclusion
on which of the possibilities discussed above is the one realised in a full computation.

Eventually we shall discuss the relation of the present approach to the Nielsen identity. As $k$ goes to zero, the effective average action only depends on one metric,
and full diffeomorphism invariance is restored. In our fluctuation language, this means that when we flow towards the \IR{}, the potential $\mathcal V$ needs
to cancel the determinant of the background metric and replace it by the determinant of the full metric in our ansatz for the effective action \eqref{eq:EHansatz}, so that
\begin{equation}
 \lim_{k\to0} \mathcal V = \Lambda_\text{IR} \sqrt{\frac{\det g}{\det \bar g}} \, ,
\end{equation}
with $\Lambda_\text{IR}$ the observed macroscopic value of the cosmological constant. This amounts to a fine-tuning problem in the \UV{} \cite{Denz:2016qks}, and can give rise to severe constraints
on the allowed trajectories in theory space, so that the number of free parameters can be less than the number of relevant directions of the fixed point. It is even feasible that
potentially viable fixed points have to be discarded if they don't admit a proper diffeomorphism invariant \IR{} regime.
For the Asymptotic Safety conjecture to work, it is necessary that diffeomorphism invariance can be combined with staying in the critical hypersurface
of the fixed point. In this way, even a fixed point with a priori infinitely many relevant directions could be physically viable, if the constraint imposed by the Nielsen identity reduces the
number of actually free parameters to a finite number.

%======================================================
\section{Summary}\label{sec:summary}

The present work laid the foundation for the study of gravitational correlation functions of arbitrary order. The central ingredient is the \CHT{}, 
which allows to rewrite many tensorial expressions in terms of scalar invariants and a small number of basis tensors. As a proof of principle, we derived and 
solved the flow equation for the graviton potential in an approximation where we retained the full dependence on conformal fluctuations and first order perturbations
by gravitational wave fluctuations. The results 
indicate strong quantum effects, emphasising the need of the present approach to reliably study the \UV{} limit. Some hints are found that there might be a 
finite graviton vacuum expectation value, or even a non-metric \UV{} phase, depending on the sign of the trace anomalous dimension. Further studies are however necessary to give a definite result. On the technical side, our approach gives a significant contribution towards the resolution of the split-Ward identity, and clarifies the relation between the bimetric and the fluctuation language.

Future studies should try to resolve the full $\mathfrak h_2$-dependence of the potential and the anomalous dimensions, together with a self-consistent flow 
equation for the Newton coupling. This allows for the self-consistent determination of critical exponents. This enhancement is technically very involved, as it needs the resolution of lots of tensor structures to 
calculate the flow equation. Nevertheless, we put forward the necessary ingredients to implement such a calculation.

Another interesting open point is the integration of a flow towards the \IR{}, which is necessary for the resolution of the Nielsen identity, and might shed light on the question of graviton condensation. The latter might however need momentum dependent invariants, \eg{} $\tr \left( \hTL \bar \Delta \hTL \right)$, which are related to the curvature of the full metric. For a related discussion 
of gravitational condensates in the Regge-Wheeler lattice formulation of quantum gravity, see \cite{Hamber:2017pli}.

\acknowledgments{The author wants to thank A.\ Eichhorn, H.\ Gies, S.\ Lippoldt, J.\ M.\ Pawlowski, F.\ Saueressig and A.\ Wipf
for useful discussions, and H.\ Gies and F.\ Saueressig
for critical comments on the manuscript.
This research was supported by the Deutsche Forschungsgemeinschaft (DFG) graduate school GRK 1523/2, by the DFG grant no. Wi 777/11
and by the Netherlands Organisation for Scientific Research (NWO) within the Foundation for Fundamental Research on Matter (FOM) grant 13VP12.
Further the author thanks the organisers of the workshop ``Quantum Spacetime and the Renormalization Group'' at Lorentz center in Leiden, where parts 
of this work have been presented and discussed.
}

%======================================================
\appendix   
%=====================================================================================================================%%=====================================================================================================================%
\section{Cayley-Hamilton theorem}
\label{sec:CHT}
Most technical results of this and the following appendices are part of the author's Ph.D. thesis \cite{dbt_mods_00032999}.
Due to the central importance of the Cayley-Hamilton theorem in this work, we shall state it here
and collect some explicit formulas for the relevant case of $4\times4$-matrices. The general theorem
can be stated as follows. Consider the characteristic polynomial $p$ of a matrix $\mathbbm A$,
\begin{equation}
 p(\lambda) = \det\left( \lambda \mathbbm{1} - \mathbbm A \right) \, .
\end{equation}
The Cayley-Hamilton theorem now states that if one replaces $\lambda$ by the matrix $\mathbbm A$ itself in
this polynomial, one gets the zero matrix,
\begin{equation}
 p(\mathbbm A) = 0 \, .
\end{equation}
Stated differently, a matrix is completely characterised by its eigenvalues,
up to similarity transformations. Moreover, the theorem provides an explicit algorithm to
convert powers of the matrix $\mathbbm A$ which are higher than its dimension to a linear combination
of lower powers of $\mathbbm A$ and the unit matrix.

Let us now specify to $4\times4$-matrices. In that case,
\small
\begin{equation}\label{eq:CHT}
 \mathbbm A^4 - (\tr \mathbbm A) \mathbbm A^3 + \frac{1}{2} \left[ (\tr \mathbbm A)^2 - \tr (\mathbbm A^2) \right] \mathbbm A^2 - \frac{1}{6} \left[ (\tr 
\mathbbm A)^3 - 3 (\tr \mathbbm A)\tr (\mathbbm A^2) + 2 \tr (\mathbbm A^3) \right] \mathbbm A + \det \mathbbm A \, \mathbbm{1}_4 = 0 \, .
\end{equation}
\normalsize
As shown in the main text, it is beneficial to divide matrices into traceless and trace parts.
Thus, let $\mathbbm A = \mathbbm B + \frac{1}{4} \tr(\mathbbm A) \mathbbm{1}_4$, where $\mathbbm B$ is the traceless part of $\mathbbm A$. Then,
\begin{equation}
 \mathbbm B^4 - \frac{1}{2} \tr (\mathbbm B^2) \mathbbm B^2 - \frac{1}{3} \tr (\mathbbm B^3) \mathbbm B + \det \mathbbm B \, \mathbbm{1}_4 = 0 \, .
\end{equation}
Any scalar function f of a constant matrix $\mathbbm A$ can thus be parameterised as $f(\tr \mathbbm A$, $\tr (\mathbbm B^2)$, $\tr (\mathbbm B^3)$, $\det 
\mathbbm B)$.
This parameterisation is useful as it allows for controlled approximations, \eg{} $\det \mathbbm B = \tr (\mathbbm B^3) = \tr (\mathbbm B^2) = 0$.
An analogous approximation with $\tr (\mathbbm A^2) = 0$ would already entail $\mathbbm A\equiv 0$ for a real and symmetric $\mathbbm A$.

\section{Parameterisation and inverse metric}
\label{sec:paramandinverse}
Let us now present the full formulas for the inverse and the determinant of the full metric in an arbitrary parameterisation. The coefficients $\mathcal B_i$ 
of \eqref{eq:invmetric_fulltext} read
\begin{align}
 \mathcal B_0 &= \frac{1}{216 \slashed\Delta} \big[ 216 \mathcal{A}_0^2 
\left(\mathcal{A}_2 \mathfrak{h}_2+\mathcal{A}_3 \mathfrak{h}_3\right)+9 
\mathcal{A}_0 \left(-12 \mathcal{A}_1^2 \mathfrak{h}_2-12 \mathcal{A}_1 \left(2 
\mathcal{A}_2
   \mathfrak{h}_3+\mathcal{A}_3 \left(\mathfrak{h}_2^2-4 
\mathfrak{h}_4\right)\right) \right. \notag \\
&\qquad\qquad \left. +4 \mathcal{A}_2 \mathcal{A}_3 
\mathfrak{h}_2 \mathfrak{h}_3+6 \mathcal{A}_2^2 \left(\mathfrak{h}_2^2+4
   \mathfrak{h}_4\right)+\mathcal{A}_3^2 \left(-3 \mathfrak{h}_2^3+24 
\mathfrak{h}_4 \mathfrak{h}_2+8 \mathfrak{h}_3^2\right)\right) \notag \\
&\qquad\qquad  +216 \mathcal{A}_2 \mathcal{A}_3^2 \mathfrak{h}_4^2+ 2
\mathfrak{h}_3 \left(4 \mathcal{A}_3^3 \mathfrak{h}_3^2+9 \mathcal{A}_1 
\left(\left(\mathcal{A}_3 \mathfrak{h}_2+2 \mathcal{A}_1\right){}^2-2 
\mathcal{A}_2^2 \mathfrak{h}_2\right) \right. \notag \\
&\qquad\qquad \left. -6 \mathcal{A}_2 \mathfrak{h}_3
   \left(\mathcal{A}_3 \left(\mathcal{A}_3 \mathfrak{h}_2+6 
\mathcal{A}_1\right)-2 \mathcal{A}_2^2\right)\right)-18 \mathfrak{h}_4 \left(3 
\mathcal{A}_2 \left(\left(\mathcal{A}_3 \mathfrak{h}_2+2
   \mathcal{A}_1\right){}^2-2 \mathcal{A}_2^2 \mathfrak{h}_2\right) \right. 
\notag \\
   &\qquad\qquad \left. +4 
\mathcal{A}_3 \mathfrak{h}_3 \left(\mathcal{A}_2^2-\mathcal{A}_3 
\left(\mathcal{A}_3 \mathfrak{h}_2+2
   \mathcal{A}_1\right)\right)\right)+216 \mathcal{A}_0^3 \big] \, , \label{eq:fullinversemetriccoeffs1} \\
 \mathcal B_1 &= \frac{1}{72 \slashed\Delta} \big[ -72 \mathcal{A}_3^3 
\mathfrak{h}_4^2+\mathfrak{h}_2 \left(4 \mathcal{A}_3^3 \mathfrak{h}_3^2+9 
\mathcal{A}_1 \left(\left(\mathcal{A}_3 \mathfrak{h}_2+2 
\mathcal{A}_1\right){}^2-2 \mathcal{A}_2^2
   \mathfrak{h}_2\right) \right. \notag \\
&\qquad\qquad \left. -6 \mathcal{A}_2 \mathfrak{h}_3 
\left(\mathcal{A}_3 \left(\mathcal{A}_3 \mathfrak{h}_2+6 \mathcal{A}_1\right)-2 
\mathcal{A}_2^2\right)\right)-18 \mathfrak{h}_4 \left(2
   \mathcal{A}_1-\mathcal{A}_3 \mathfrak{h}_2\right) \times \notag \\
&\qquad\qquad \left(\mathcal{A}_3 \left(\mathcal{A}_3 \mathfrak{h}_2+2 
\mathcal{A}_1\right)-2 \mathcal{A}_2^2\right)-12 \mathcal{A}_0 
\left(-\mathfrak{h}_3
   \left(\mathcal{A}_3^2 \mathfrak{h}_2+2 \mathcal{A}_2^2\right) \right. \notag 
\\
&\qquad\qquad \left. +2 \mathcal{A}_1 \left(3 \mathcal{A}_2 
\mathfrak{h}_2+\mathcal{A}_3 \mathfrak{h}_3\right)+12 \mathcal{A}_2 \mathcal{A}_3
   \mathfrak{h}_4\right)-72 \mathcal{A}_0^2 \mathcal{A}_1 \big] \, , \\
 \mathcal B_2 &= \frac{1}{216 \slashed\Delta} \big[ -18 \mathcal{A}_0 \left(-3 
\left(\mathcal{A}_3 \mathfrak{h}_2+2 \mathcal{A}_1\right){}^2+6 \mathcal{A}_2 
\left(\mathcal{A}_2 \mathfrak{h}_2+2 \mathcal{A}_0\right)+4 \mathcal{A}_2 
\mathcal{A}_3
   \mathfrak{h}_3\right) \notag \\
&\qquad\qquad -36 \mathfrak{h}_4 \left(-3 \mathcal{A}_3 
\mathcal{A}_2 \left(\mathcal{A}_3 \mathfrak{h}_2+4 \mathcal{A}_1\right)+2 
\mathcal{A}_3^2 \left(\mathcal{A}_3 \mathfrak{h}_3+3
   \mathcal{A}_0\right)+6 \mathcal{A}_2^3\right) \big] \, , \\
 \mathcal B_3 &= \frac{1}{36 \slashed\Delta} \big[ -4 \mathcal{A}_3^3 
\mathfrak{h}_3^2-18 \mathcal{A}_3 \mathfrak{h}_4 \left(\mathcal{A}_3 
\left(\mathcal{A}_3 \mathfrak{h}_2+2 \mathcal{A}_1\right)-2 
\mathcal{A}_2^2\right) \notag \\
&\qquad\qquad -9 \left(4
   \mathcal{A}_3 \mathcal{A}_1^2 \mathfrak{h}_2+\mathcal{A}_1 \left(\mathfrak{h}_2 
\left(\mathcal{A}_3^2 \mathfrak{h}_2-2 \mathcal{A}_2^2\right)-8 \mathcal{A}_0 
\mathcal{A}_2\right)+4 \mathcal{A}_1^3+4
   \mathcal{A}_0^2 \mathcal{A}_3\right) \notag \\ 
&\qquad\qquad +6 \mathfrak{h}_3 \left(\mathcal{A}_3 
\mathcal{A}_2 \left(\mathcal{A}_3 \mathfrak{h}_2+6 \mathcal{A}_1\right)-2 
\mathcal{A}_2^3-4 \mathcal{A}_0
   \mathcal{A}_3^2\right) \big] \, , \label{eq:fullinversemetriccoeffs4}
\end{align}
where we already used the ratio of determinants
\begin{align}
 \slashed\Delta \equiv \frac{\det g}{\det \bar g} &= \mathcal{A}_0^3 
\left(\mathcal{A}_2 \mathfrak{h}_2+\mathcal{A}_3 
\mathfrak{h}_3\right)+\frac{1}{24} \mathcal{A}_0^2 \left(-12 \mathcal{A}_1^2 
\mathfrak{h}_2-12 \mathcal{A}_1 \left(2 \mathcal{A}_2
   \mathfrak{h}_3+\mathcal{A}_3 \left(\mathfrak{h}_2^2-8 
\mathfrak{h}_4\right)\right) \right. \notag \\
&\quad \left. +4 \mathcal{A}_2 \mathcal{A}_3 
\mathfrak{h}_2 \mathfrak{h}_3+6 \mathcal{A}_2^2 \left(\mathfrak{h}_2^2+8
   \mathfrak{h}_4\right)+\mathcal{A}_3^2 \left(-3 \mathfrak{h}_2^3+36 
\mathfrak{h}_4 \mathfrak{h}_2+8 \mathfrak{h}_3^2\right)\right) \notag \\
&\quad+\frac{1}{108} 
\mathcal{A}_0 \left(432 \mathcal{A}_2 \mathcal{A}_3^2
   \mathfrak{h}_4^2-18 \mathfrak{h}_4 \left(3 \mathcal{A}_2 \left(\mathcal{A}_3^2 
\mathfrak{h}_2^2-2 \left(\mathcal{A}_2^2-2 \mathcal{A}_1 \mathcal{A}_3\right) 
\mathfrak{h}_2+8
   \mathcal{A}_1^2\right) \right.\right. \notag \\
   &\quad\left.+\mathcal{A}_3 \mathfrak{h}_3 \left(2 
\mathcal{A}_2^2-\mathcal{A}_3 \left(3 \mathcal{A}_3 \mathfrak{h}_2+10 
\mathcal{A}_1\right)\right)\right) \notag \\
&\quad +\mathfrak{h}_3 \left(4
   \mathcal{A}_3^3 \mathfrak{h}_3^2+9 \mathcal{A}_1 \left(\left(\mathcal{A}_3 
\mathfrak{h}_2+2 \mathcal{A}_1\right){}^2-2 \mathcal{A}_2^2 
\mathfrak{h}_2\right) \right. \notag \\
&\quad \left.\left.-6 \mathcal{A}_2 \mathfrak{h}_3
   \left(\mathcal{A}_3 \left(\mathcal{A}_3 \mathfrak{h}_2+6 
\mathcal{A}_1\right)-2 \mathcal{A}_2^2\right)\right)\right) \notag \\
&\quad +\frac{1}{36} 
\mathfrak{h}_4 \left(36 \mathcal{A}_3^4
   \mathfrak{h}_4^2+\mathcal{A}_1 \left(4 \mathcal{A}_3^3 \mathfrak{h}_3^2+9 
\mathcal{A}_1 \left(\left(\mathcal{A}_3 \mathfrak{h}_2+2 
\mathcal{A}_1\right){}^2-2 \mathcal{A}_2^2 \mathfrak{h}_2\right) \right.\right. \notag 
\\
&\quad\left.-6
   \mathcal{A}_2 \mathfrak{h}_3 \left(\mathcal{A}_3 \left(\mathcal{A}_3 
\mathfrak{h}_2+6 \mathcal{A}_1\right)-2 \mathcal{A}_2^2\right)\right)+6 
\mathfrak{h}_4 \left(-3 \mathcal{A}_3 \mathcal{A}_2^2
   \left(\mathcal{A}_3 \mathfrak{h}_2+8 \mathcal{A}_1\right) \right. \notag \\
   &\quad  \left.\left.+2 
\mathcal{A}_3^3 \mathcal{A}_2 \mathfrak{h}_3+6 \mathcal{A}_1 \mathcal{A}_3^2 
\left(\mathcal{A}_3 \mathfrak{h}_2+2 \mathcal{A}_1\right)+6
   \mathcal{A}_2^4\right)\right)+\mathcal{A}_0^4 \, .
\end{align}
We shall also derive the exact representation of the exponential parameterisation. For this, we first derive a recursion for the $n$-th power of $\hTL$ for 
$n\geq4$. Making the ansatz
\begin{equation}
 \left[ \hTL \right]^n = a_n \mathbbm 1_4 + b_n \hTL + c_n \left[ \hTL \right]^2 + d_n \left[ \hTL \right]^3 \, ,
\end{equation}
and with the initial conditions from the \CHT{},
\begin{equation}
 a_4 = -\mathfrak h_4 \, , \quad b_4 = \frac{1}{3} \mathfrak h_3 \, , \quad c_4 
= \frac{1}{2} \mathfrak h_2 \, , \quad d_4 = 0 \, ,
\end{equation}
we obtain the recursion
\begin{equation}\label{eq:CHT_recrel}
\begin{aligned}
 a_{n+1} &= - \mathfrak h_4 d_n \, , \\
 b_{n+1} &= a_n + \frac{1}{3} \mathfrak h_3 d_n \, , \\
 c_{n+1} &= b_n + \frac{1}{2} \mathfrak h_2 d_n \, , \\
 d_{n+1} &= c_n \, .
\end{aligned}
\end{equation}
This recursion can be solved by Mathematica, and we shall not present the result for arbitrary $n$. Rather, focusing on the exponential parameterisation, one 
can transform this set of recursion relations to a set of differential equations in a fiducial variable $x$. For this, we introduce the functions
\begin{equation}
 A(x) = \sum_{n=4}^\infty \frac{a_n}{n!} x^n \, , \quad B(x) = \sum_{n=4}^\infty \frac{b_n}{n!} x^n \, , \quad C(x) = \sum_{n=4}^\infty \frac{c_n}{n!} x^n \, , 
\quad D(x) = \sum_{n=4}^\infty \frac{d_n}{n!} x^n \, .
\end{equation}
The metric in exponential parameterisation then reads
\begin{equation}
\begin{aligned}
 g^\text{exp} &= \bar g \, e^{\mathfrak h_1/4}\, \times \\
 & \quad\left[ \mathbbm 1_4 (1 + A(1)) + \hTL \, (1 + B(1)) + \left[ \hTL \right]^2 \left( 
\frac{1}{2} + C(1) \right) + \left[ \hTL \right]^3 \left( \frac{1}{6} + D(1) \right) \right] \, .
\end{aligned}
\end{equation}
The functions $A, B, C, D$ are the solutions to the set of ordinary differential equations obtained by multiplying the recursion relations with 
$\tfrac{x^n}{n!}$ and summing over $n$ from $4$ to $\infty$. Doing so, one 
arrives at
\begin{equation}
\begin{aligned}
 A'(x) &= \mathfrak h_4 \left( \frac{x^3}{6} - D(x) \right) \, , \\
 B'(x) &= A(x) + \frac{1}{3} \mathfrak h_3 \left( \frac{x^3}{6} + D(x) \right) 
\, , \\
 C'(x) &= B(x) + \frac{1}{2} \mathfrak h_2 \left( \frac{x^3}{6} + D(x) \right) 
\, , \\
 D'(x) &= C(x) \, .
\end{aligned}
\end{equation}
Initial conditions for the functions at $x=0$ follow from their definition. This set of differential equations can be solved by Mathematica. To present the 
result, we first introduce the polynomial
\begin{equation}
 p(y) = 6\mathfrak h_4 - 2 \mathfrak h_3 y - 3 \mathfrak h_2 y^2 + 6 y^4 \, ,
\end{equation}
and define the operator $\mathcal{RS}$ (for \textit{RootSum}) which maps a function to the sum of the values of this function at the roots of $p$,
\begin{equation}
 \mathcal{RS}[f(y)] = \sum_{y_i : p(y_i)=0} f(y_i) \, .
\end{equation}
With a final abbreviation,
\begin{equation}
 \rho_n = \mathcal{RS} \left[ \frac{e^y y^n}{-\mathfrak h_3 - 3\mathfrak h_2 y 
+ 12y^3} \right] \, ,
\end{equation}
we find for the functions at $x=1$,
\begin{align}
 A(1) &= \mathcal{RS} \left[ \frac{1}{12 y^4 \left(-\mathfrak{h}_3+12 y^3-3 
\mathfrak{h}_2 y\right)} \left( \mathfrak{h}_4 e^{-y} \left(y (y (y+3)+6)-6 
e^y+6\right) \times \right. \right. \notag \\
&\qquad\qquad \left(2 \left(2 \mathfrak{h}_3^2 \rho _0-9 \rho 
_3 y^3+9 
\mathfrak{h}_4 \left(\rho _2+y \left(\rho _1+\rho _0 y\right)\right)+3
   \mathfrak{h}_3 \left(2 \rho _3+y \left(\rho _2+\rho _1 
y\right)\right)\right) \right. \notag \\
&\qquad\qquad \left.\left.\left. -9 \mathfrak{h}_2^2 \rho _1 y+3 
\mathfrak{h}_2 \left(-6 
\mathfrak{h}_4 \rho _0+2 \mathfrak{h}_3 \left(\rho _0 y-\rho
   _1\right)+3 y \left(2 \rho _3+y \left(\rho _2+2 \rho _1 
y\right)\right)\right)\right) \right) \right] \, , \\
 B(1) &= \mathcal{RS} \left[ \frac{1}{12 y^4 \left(-\mathfrak{h}_3+12 y^3-3 
\mathfrak{h}_2 y\right)} \left( e^{-y} \left(y (y (y+3)+6)-6 e^y+6\right) 
\times \right.\right. \notag \\
&\qquad\qquad \left(3 \mathfrak{h}_2 \left(\mathfrak{h}_4 
\left(\rho _0 \left(6 y^3-2 
\mathfrak{h}_3\right)+6 \left(\rho _3+y \left(\rho _2+\rho _1
   y\right)\right)\right)-\mathfrak{h}_3 \rho _2 y^2\right) \right. \notag \\
&\qquad\qquad +2 \left(9 
\mathfrak{h}_4 \left(\mathfrak{h}_4 \left(\rho _1+\rho _0 y\right)-y^2 
\left(\rho _3+\rho _2 y\right)\right)+3 \mathfrak{h}_3
   \left(\mathfrak{h}_4 \left(\rho _2+\rho _0 y^2\right)-\rho _3 
y^3\right) \right. \notag \\
&\qquad\qquad \left.\left.\left.\left. -\mathfrak{h}_3^2 y \left(\rho _2+\rho 
_1 y\right)\right)-9 
\mathfrak{h}_4 \mathfrak{h}_2^2 \left(\rho _1+\rho _0
   y\right)\right) \right) \right] \, , \\
 C(1) &= \mathcal{RS} \left[ -\frac{1}{24 y^4 
\left(-\mathfrak{h}_3+12 y^3-3 \mathfrak{h}_2 y\right)} \left( e^{-y} \left(y (y 
(y+3)+6)-6 e^y+6\right) \times \right.\right. \notag \\
&\qquad\qquad \left(9 \mathfrak{h}_2^2 \rho _2 y^2+4 \left(3 
\mathfrak{h}_3 \left(y^2 
\left(\rho _3+\rho _2 y\right)-\mathfrak{h}_4 \left(\rho _1+\rho _0
   y\right)\right)+\mathfrak{h}_3^2 \rho _1 y \right.\right. \notag \\
&\qquad\qquad \left. +9 \mathfrak{h}_4 
\left(y 
\left(\rho _3+y \left(\rho _2+\rho _1 y\right)\right)-\mathfrak{h}_4 \rho 
_0\right)\right) \notag \\
&\qquad\qquad \left.\left.\left. +6 \mathfrak{h}_2 \left(3 \rho _3
   y^3-3 \mathfrak{h}_4 \left(\rho _2+\rho _0 y^2+2 \rho _1 
y\right)+\mathfrak{h}_3 y \left(\rho _2+\rho _1 y\right)\right)\right) \right) 
\right] \, , \\
 D(1) &= \mathcal{RS} \left[ -\frac{1}{4 y^4 \left(-\mathfrak{h}_3+12 y^3-3 
\mathfrak{h}_2 y\right)} \left( e^{-y} \left(y (y (y+3)+6)-6 e^y+6\right) 
\times \right.\right. \notag \\
&\qquad\qquad \left(\mathfrak{h}_4 \left(\rho _0 \left(-4 
\mathfrak{h}_3+6 y^3-6 
\mathfrak{h}_2 y\right)+6 \left(\rho _3+\rho _1
   \left(y^2-\mathfrak{h}_2\right)+\rho _2 y\right)\right) \right. \notag \\
&\qquad\qquad \left.\left.\left.  +y \left(3 
\mathfrak{h}_2 y \left(\rho _3+\rho _2 y\right)+2 \mathfrak{h}_3 \left(\rho 
_3+y 
\left(\rho _2+\rho _1
   y\right)\right)\right)\right) \right) \right] \, .
\end{align}
For the special case where we neglect the invariants $\mathfrak h_3$ and 
$\mathfrak h_4$, we find $A=B=\mathcal O(\mathfrak h_3, \mathfrak h_4)$ and
\begin{align}
 C(1) &= -\frac{4+\mathfrak h_2 - 4 \cosh \sqrt{\frac{\mathfrak 
h_2}{2}}}{2\mathfrak h_2} + \mathcal O(\mathfrak h_3, \mathfrak h_4) \, , \\
 D(1) &= -\frac{1}{6} - \frac{2}{\mathfrak h_2} + \frac{2\sqrt{2} \sinh 
\sqrt{\frac{\mathfrak h_2}{2}}}{\mathfrak h_2^{3/2}} + \mathcal O(\mathfrak h_3, 
\mathfrak h_4) \, .
\end{align}
These expressions admit a Taylor expansion in $\mathfrak h_2$ around zero and are thus regular also for $\mathfrak h_2\to0$.

\section{Curvature identities}
\label{sec:RtoChristoffelsquared}
In this appendix, we present some useful formulas related to the curvature tensors.
First, we rewrite the kinetic part of the Einstein-Hilbert action
into a form of which the second derivative with respect to the fluctuation simplifies tremendously.
The Ricci scalar can be expressed as
\begin{equation}
 R = g^{\alpha\beta} \left[ {\Gamma^\gamma}_{\gamma\delta} {\Gamma^\delta}_{\alpha\beta} - {\Gamma^\gamma}_{\alpha\delta} {\Gamma^\delta}_{\gamma\beta}
 -\partial_\alpha {\Gamma^\gamma}_{\gamma\beta} + \partial_\gamma {\Gamma^\gamma}_{\alpha\beta} \right] \, ,
\end{equation}
where $\Gamma$ is the Christoffel symbol of the metric $g$, and $\partial$ denotes the standard partial derivative.
Using basic identities from differential geometry, partial integration and dropping boundary terms, we can rewrite
\begin{equation}
\begin{aligned}
 & \int \text{d}^d x \sqrt{\det g} \, g^{\alpha\beta} \left[ -\partial_\alpha {\Gamma^\gamma}_{\alpha\beta} + \partial_\gamma {\Gamma^\gamma}_{\alpha\beta} \right] \\
 = & \int \text{d}^d x \left[ {\Gamma^\gamma}_{\gamma\beta} \, \partial_\alpha \left( \sqrt{\det g} \, g^{\alpha\beta} \right) - {\Gamma^\gamma}_{\alpha\beta} \, \partial_\gamma \left( \sqrt{\det g} \, g^{\alpha\beta} \right) \right] \\
 = & \int \text{d}^d x \left[ {\Gamma^\gamma}_{\gamma\beta} \left( -\sqrt{\det g} \, g^{\mu\nu} {\Gamma^\beta}_{\mu\nu} \right) - {\Gamma^\gamma}_{\alpha\beta} \sqrt{\det g} \left({\Gamma^\delta}_{\gamma\delta}  g^{\alpha\beta} - \left( {\Gamma^\alpha}_{\gamma\delta} g^{\delta\beta} + {\Gamma^\beta}_{\gamma\delta} g^{\alpha\delta} \right) \right) \right] \\
 = & \int \text{d}^d x \sqrt{\det g} \, g^{\alpha\beta} \left[ - 2 {\Gamma^\gamma}_{\gamma\delta} {\Gamma^\delta}_{\alpha\beta} + 2 {\Gamma^\gamma}_{\alpha\delta} {\Gamma^\delta}_{\gamma\beta} \right] \, ,
\end{aligned}
\end{equation}
and thus
\begin{equation}
 \int \text{d}^d x \sqrt{\det g} \, R = \int \text{d}^d x \sqrt{\det g} \, g^{\alpha\beta} \left[ - {\Gamma^\gamma}_{\gamma\delta} {\Gamma^\delta}_{\alpha\beta} + {\Gamma^\gamma}_{\alpha\delta} {\Gamma^\delta}_{\gamma\beta} \right] \, .
\end{equation}
The virtue of this rewriting is the following: for the flow equation, we need the second variation of the action with respect to the fluctuation.
In the present work, we project the flow equation onto constant $h$ and flat background $\bar g = \delta$. Since the Christoffel symbols are linear in derivatives,
the only contribution to the second variation comes from the combination where all Christoffel symbols are varied. As a side remark, the same result can be obtained if one treats
the Christoffel symbols as $(1,2)$-tensors and rewrites the partial derivatives as covariant derivatives plus the corresponding Christoffel symbols, finally 
dropping the terms with covariant derivatives.

For future reference, we also write formulas for the Ricci scalar, Ricci tensor and Riemann tensor in terms of the background curvatures and derivatives in a 
background-covariant way. A simple calculation shows that
\begin{equation}
\begin{aligned}
 R_{\mu\nu} &= \bar R_{\mu\nu} + \frac{1}{2} g^{\alpha\beta} \left( 2 \bar D_\beta \bar D_{(\mu} g_{\nu)\alpha} - \bar 
D_\alpha \bar D_\beta g_{\mu\nu} - \bar D_\mu \bar D_\nu g_{\alpha\beta} \right) \\
& + \frac{1}{4} g^{\alpha\beta} g^{\gamma\delta} \Big[ \left( 2\bar D_\beta g_{\gamma\delta} \bar D_{(\mu} g_{\nu)\alpha} 
+ \bar D_\mu g_{\alpha\gamma} \bar D_\nu g_{\beta\delta} - \bar D_\alpha g_{\mu\nu} \bar D_\beta g_{\gamma\delta} \right) \\
& + 2 \left( \bar D_\alpha g_{\mu\nu} \bar D_\delta g_{\beta\gamma} + 2\bar D_\gamma g_{\mu\alpha} \bar D_{[\delta} g_{\beta]\nu}
 - 2\bar D_\delta g_{\beta\gamma} \bar D_{(\mu} g_{\nu)\alpha} \right) \Big] \, ,
\end{aligned}
\end{equation}
from which we immediately get the Ricci scalar by a contraction with $g^{\mu\nu}$. For the Riemann tensor, we have
\begin{align}
%  R_{\mu\nu\rho\sigma} &= g_{\alpha[\mu} \bar R_{\nu]}{}^\alpha{}_{\rho\sigma} - \bar D_\rho \bar D_{[\mu} g_{\nu]\sigma} + \bar D_\sigma \bar D_{[\mu} 
% g_{\nu]\rho} \notag \\
% &+ \frac{1}{4} g^{\alpha\beta} \Big[ \bar D_\alpha g_{\mu\sigma} \bar D_\beta g_{\nu\rho} - \bar D_\alpha g_{\mu\rho} \bar D_\beta g_{\nu\sigma} + \bar 
% D_\alpha g_{\nu\sigma} \bar D_\mu g_{\rho\beta} - \bar D_\alpha g_{\nu\rho} \bar D_\mu g_{\sigma\beta} \notag \\
% &\qquad\qquad +\bar D_\mu g_{\sigma\beta} \bar D_\nu g_{\rho\alpha} - \bar D_\alpha g_{\mu\sigma} \bar D_{\nu} g_{\rho\beta} + \bar D_\alpha g_{\mu\rho} \bar 
% D_\nu g_{\sigma\beta} - \bar D_\mu g_{\rho\alpha} \bar D_\nu g_{\sigma\beta} \notag \\
% &\qquad\qquad + \bar D_\beta g_{\nu\sigma} \bar D_\rho g_{\mu\alpha} - \bar D_\nu g_{\sigma\beta} \bar D_\rho g_{\mu\alpha} + \bar D_\mu g_{\sigma\beta} \bar 
% D_\rho g_{\nu\alpha} - \bar D_\alpha g_{\mu\sigma} \bar D_\rho g_{\nu\beta} \notag \\
% &\qquad\qquad - \bar D_\beta g_{\nu\rho} \bar D_\sigma g_{\mu\alpha} + \bar D_\nu g_{\rho\beta} \bar D_\sigma g_{\mu\alpha} + \bar D_\rho g_{\nu\beta} \bar 
% D_\sigma g_{\mu\alpha} - \bar D_\mu g_{\rho\beta} \bar D_\sigma g_{\nu\alpha} \notag \\ 
% &\qquad\qquad + \bar D_\alpha g_{\mu\rho} \bar D_\sigma g_{\nu\beta} - \bar D_\rho g_{\mu\alpha} \bar D_\sigma g_{\nu\beta} \Big] \, .
 R_{\mu\nu\rho\sigma} &= g_{\alpha[\mu} \bar R_{\nu]}{}^\alpha{}_{\rho\sigma} - \bar D_\rho \bar D_{[\mu} g_{\nu]\sigma} + \bar D_\sigma \bar D_{[\mu} 
g_{\nu]\rho} \notag \\
&+ \frac{1}{2} g^{\alpha\beta} \Big[ \bar D_\alpha g_{\mu[\sigma} \bar D_{|\beta|} g_{\rho]\nu} + \bar D_\alpha g_{\nu[\sigma} \bar D_{|\mu|} g_{\rho]\beta}
+ \bar D_{[\mu} g_{|\sigma\beta|} \bar D_{\nu]} g_{\rho\alpha} + \bar D_\alpha g_{\mu[\rho} \bar D_{|\nu|} g_{\sigma]\beta} \notag \\
&\mkern-9.5mu + \bar D_\beta g_{\nu[\sigma} \bar D_{\rho]} g_{\mu\alpha} + \bar D_\mu g_{\beta[\sigma} \bar D_{\rho]} g_{\nu\alpha}
+ \bar D_\nu g_{\beta[\rho} \bar D_{\sigma]} g_{\mu\alpha} + \bar D_{[\rho} g_{|\nu\beta|} \bar D_{\sigma]} g_{\mu\alpha}
+ \bar D_\alpha g_{\mu[\rho} \bar D_{\sigma]} g_{\nu\beta} \Big] \, .
\end{align}

\section{Quantum gauge transformation}\label{sec:gf}

In this appendix we discuss some aspects of our choice of gauge fixing. In particular, we choose to gauge-fix the fluctuation field $h$ instead of the 
full metric $g$, since then the vertices are not affected by the gauge fixing, which might yield results that are less sensitive to the specific gauge choice. 
For the corresponding ghost action, we have to derive how the ``quantum gauge transformation'' of $h$ looks like in 
terms of the variation of $g$ \cite{Nink:2014yya}. The latter is nothing else than a BRST transformation along the ghost field $c$,
\begin{equation}
 \delta_Q g_{\mu\nu} = \mathcal L_c g_{\mu\nu} = D_\mu c_\nu + D_\nu c_\mu \, ,
\end{equation}
$\mathcal L$ being the Lie derivative.
We will show that this can be done practically in all generality, again with the help of the \CHT{}, and 
also derive the explicit expression that we need in the main text.

To illustrate the general calculation, we start with \eqref{eq:metric_fulltext}, where we however do not use a traceless decomposition,
\begin{equation}
 g = \bar g \left( \tilde{\mathcal A}_0 \, \mathbbm 1_4  + \tilde{\mathcal A}_1 \, \hmat + \tilde{\mathcal A}_2 \, \hmat^2 + \tilde{\mathcal A}_3 \,
\hmat^3 \right) \, .
\end{equation}
The $\tilde{\mathcal A}_i$ are understood depend on the traces and the determinant of $\hmat$, \ie{} $\tilde{\mathcal A}_i = \tilde{\mathcal A}_i(\tr \hmat, \tr \hmat^2, \tr \hmat^3, \det \hmat)$.
A quantum gauge transformation of this equation gives (remember that this means $\delta_Q \bar g=0$)
\begin{equation}\label{eq:variedsplit}
\begin{aligned}
 \delta_Q g &= \bar g \Big( (\delta_Q \tilde{\mathcal A}_0) \, \mathbbm 1_4 + (\delta_Q \tilde{\mathcal A}_1) \hmat + \tilde{\mathcal A}_1 
\, \delta_Q\hmat + (\delta_Q \tilde{\mathcal A}_2) \, \hmat^2 + \tilde{\mathcal A}_2 \big[ (\delta_Q \hmat) \hmat + \hmat \delta_Q \hmat \big] \\
&\qquad+ (\delta_Q \tilde{\mathcal A}_3) \, \hmat^3 + \tilde{\mathcal A}_3 \big[ (\delta_Q\hmat)\hmat^2 + \hmat(\delta_Q\hmat)\hmat + \hmat^2\delta_Q\hmat \big] \Big) 
\, ,
\end{aligned}
\end{equation}
with
\begin{equation}
\begin{aligned}
 \delta_Q\tilde{\mathcal A}_i &= \tilde{\mathcal A}_i^{(1,0,0,0)} \tr(\delta_Q\hmat) + 2 \tilde{\mathcal A}_i^{(0,1,0,0)} \tr((\delta_Q\hmat)\hmat) + 
3\tilde{\mathcal A}_i^{(0,0,1,0)} \tr((\delta_Q\hmat)\hmat^2) \\
&\quad + \frac{1}{24} \tilde{\mathcal A}_i^{(0,0,0,1)} \Big[4 (\tr\hmat)^3 \tr(\delta_Q\hmat) - 12 (\tr(\hmat) \tr(\hmat^2) \tr(\delta_Q\hmat) + (\tr\hmat)^2 
\tr(\hmat\delta_Q\hmat)) \\
&\qquad + 12 \tr(\hmat^2) \tr(\hmat\delta_Q\hmat) + 8 \tr(\hmat^3) \tr(\delta_Q\hmat) + 24 \tr(\hmat) \tr(\hmat^2\delta_Q\hmat) - 
24\tr(\hmat^3\delta_Q\hmat) \Big] \, .
\end{aligned}
\end{equation}
The superscripts indicate the number of derivatives w.r.t.\ the respective arguments, \ie{} $\tilde{\mathcal A}_i^{(1,0,0,0)} = \partial_{\tr\hmat}\tilde{\mathcal A}_i$ and so on.
The task is to solve \eqref{eq:variedsplit} for $\delta_Q\hmat$. Clearly, $\delta_Q\hmat$ will be linear in $\delta_Q g$, and in general contains all 
possible products of $\delta_Q g$ with $\hmat$. Due to the \CHT{}, there is only a finite number of independent products, and in four dimensions, this number is 
32. By making an ansatz for $\delta_Q\hmat$ as linear combination of the elements of these products, inserting this ansatz into \eqref{eq:variedsplit}, and 
using the \CHT{}, we can solve for the coefficients. We calculated the solution explicitly with the command \textit{SolveConstants} of xTras 
\cite{2014CoPhC.185.1719N}, but the result is too bulky to be presented here. Recently, an order-by-order calculation has been discussed in \cite{Eichhorn:2017sok}.

It is obvious that in a similar fashion we can derive the relation between the variation of $h$ and the variation of the background
metric, $\delta_B\bar g$, with fixed variation of the full metric, $\delta_B g=0$. This corresponds to a background gauge transformation, and plays a role
in the construction of explicit solutions to the split-Ward identity, see \eg{} \cite{Percacci:2016arh}.

Let us now discuss the exponential split in particular, and derive the quantum transformation of $h$ to quadratic order in $\hTL$, including all information on 
the trace. For this, we start with the definition of the exponential split,
\begin{equation}
 g_{\mu\nu} = \bar g_{\mu\rho} e^{\mathfrak h_1/4} \left[ e^{\hTL} \right]^\rho{}_\nu = \bar g_{\mu\rho} e^{\mathfrak h_1/4} \sum_{n=0}^\infty \frac{1}{n!} 
\left[ \left( \hTL{} \right)^n \right]^\rho{}_\nu \, ,
\end{equation}
and take a quantum variation with fixed background metric,
\begin{equation}
\begin{aligned}
 \delta_Q g_{\mu\nu} &= \bar g_{\mu\rho} e^{\mathfrak h_1/4} \left[ \frac{1}{4} \delta_Q \mathfrak h_1 \sum_{n=0}^\infty \frac{1}{n!} 
\left[ \left( \hTL{} \right)^n \right]^\rho{}_\nu + \sum_{n=0}^\infty \frac{1}{n!} \delta_Q \left[ \left( \hTL{} \right)^n \right]^\rho{}_\nu \right] \\
&\mkern-18mu= \bar g_{\mu\rho} e^{\mathfrak h_1/4} \left[ \frac{1}{4} \delta_Q \mathfrak h_1 \sum_{n=0}^\infty \frac{1}{n!} 
\left[ \left( \hTL{} \right)^n \right]^\rho{}_\nu + \sum_{n=0}^\infty \frac{1}{n!} \sum_{l=0}^{n-1} \left[ \left( \hTL{} \right)^l \left(\delta_Q\hTL\right) \left( \hTL{} \right)^{n-l-1} \right]^\rho{}_\nu \right] \, .
\end{aligned}
\end{equation}
Now we use that, by definition, $\delta_Q \hTL = \delta_Q \hmat - \frac{1}{4} \mathbbm 1_4 \delta_Q \mathfrak h_1$, and we see that the part including 
$\delta_Q \mathfrak h_1$ exactly cancels, leaving us with
\begin{equation}
 \delta_Q g_{\mu\nu} = \bar g_{\mu\rho} e^{\mathfrak h_1/4} \sum_{n=0}^\infty \frac{1}{n!} \sum_{l=0}^{n-1} \left[ \left( \hTL{} \right)^l \left(\delta_Q \hmat\right) \, \left( \hTL{} \right)^{n-l-1} 
\right]^\rho{}_\nu \, .
\end{equation}
We can further reorganise the sums to yield the final expression
\begin{equation}
 \delta_Q g_{\mu\nu} = \bar g_{\mu\rho} e^{\mathfrak h_1/4} \sum_{l=0}^\infty \sum_{m=0}^\infty \frac{1}{(l+m+1)!} \left[ \left( \hTL{} \right)^l \left(\delta_Q \hmat\right) \, \left( \hTL{} \right)^{m} 
\right]^\rho{}_\nu \, .
\end{equation}
For the truncation in the main text, we need $\delta_Q h$ including up to second order in $\hTL$. It is easy to show that the solution is
\begin{equation}\label{eq:qu_ga_tr_exp}
\begin{aligned}
 e^{\mathfrak h_1/4} \delta_Q h_{\mu\nu} &= \delta_Q g_{\mu\nu} - \frac{1}{2} \left( h^\text{TL}_{\mu\alpha} \bar g^{\alpha\beta} \delta_Q g_{\beta\nu} + 
\left(\delta_Q g_{\mu\alpha}\right) \bar g^{\alpha\beta} h^\text{TL}_{\beta\nu} \right) \\
&\mkern-18mu+ \frac{1}{12} \left( h^\text{TL}_{\mu\alpha} \bar g^{\alpha\beta} h^\text{TL}_{\beta\gamma} \bar g^{\gamma\sigma} \delta_Q g_{\sigma\nu} + 4 
h^\text{TL}_{\mu\alpha} \bar g^{\alpha\beta} \left(\delta_Q g_{\beta\gamma}\right) \bar g^{\gamma\sigma} h^\text{TL}_{\sigma\nu} + \left(\delta_Q g_{\mu\alpha}\right) \bar g^{\alpha\beta} 
h^\text{TL}_{\beta\gamma} \bar g^{\gamma\sigma} h^\text{TL}_{\sigma\nu} \right) \\
&\mkern-18mu+ \mathcal O(h^\text{TL}{}^3) \, .
\end{aligned}
\end{equation}

\section{Loop momentum integration}\label{sec:loopint}

In this appendix, we discuss some technical aspects in how to treat the integrals over loop momenta. Most of the discussion will be done in arbitrary 
dimension $d$ and in flat space. For the discussion, let us introduce $\mathcal P= p_\mu {T^\mu}_\nu p^\nu$, where $T$ is an arbitrary tensor and $p$ is the loop momentum. The 
first step is to calculate expressions of the type
\begin{equation}
 \int \text{d}^dp \, g(p) \, \mathcal P^n \, ,
\end{equation}
for general functions $g(p)$ which shall only depend on the absolute value of $p$. Inserting the definition of $\mathcal P$, we need to calculate
\begin{equation}
 \int \text{d}^d p \, g(p)\,p_{\mu_1} \cdots p_{\mu_{2n}} \, .
\end{equation}
By Lorentz invariance, the tensor structure (in a flat space) is given by the symmetrised product of $n$ metrics,
\begin{equation}
 \int \text{d}^d p \, g(p)\,p_{\mu_1} \cdots p_{\mu_{2n}} = \alpha_n \eta_{(\mu_1 \mu_2} \cdots \eta_{\mu_{2n-1}\mu_{2n})} \int \text{d}^d p \, g(p) \, 
p^{2n} \, .
\end{equation}
To calculate the constants $\alpha_n$, we multiply this equation by another product of $n$ metrics; the result is
\begin{equation}
 \alpha_n = \frac{(2n-1)!! (d-2)!!}{(d+2(n-1))!!} \, .
\end{equation}
Now, we have to carry out the contraction of the metrics with the product of $T$s. It is straightforward to show that
\begin{equation}
 T^{\mu_1\mu_2} \cdots T^{\mu_{2n-1}\mu_{2n}} \eta_{(\mu_1 \mu_2} \cdots \eta_{\mu_{2n-1}\mu_{2n})} = \frac{2^n}{(2n-1)!!} B_n\left( \frac{0!}{2} \tau_1, 
\frac{1!}{2} \tau_2, \dots, \frac{(n-1)!}{2} \tau_n \right) \, .
\end{equation}
Here, we introduced the traces $\tau_i = \tr \left( T^i \right)$, and $B_n$ stands for the $n$-th complete Bell polynomial.
The complete Bell polynomials are defined by
\begin{equation}
 B_n(x_1,\dots,x_n) = \sum_{k=1}^n \sum \frac{n!}{j_1! \dots j_{n-k+1}!} \left( \frac{x_1}{1!} \right)^{j_1} \dots \left( \frac{x_{n-k+1}}{(n-k+1)!} \right)^{j_{n-k+1}} \, ,
\end{equation}
where the inner sum extends over all non-negative integers $j_l$ subject to the two conditions
\begin{equation}
 \sum_{l=1}^{n-k+1} j_l = k \, , \quad \sum_{l=1}^{n-k+1} l \, j_l = n \, .
\end{equation}
Combining this, one gets
\begin{equation}
 \int \text{d}^d p \, g(p) \, \left( p_\mu p_\nu T^{\mu\nu} \right)^n = \frac{2^n (d-2)!!}{(d+2(n-1))!!} B_n \int \text{d}^d p \, g(p) \, p^{2n} = 
\frac{1}{\left( \frac{d}{2} \right)_n} B_n \int \text{d}^d p \, g(p) \, p^{2n} \, .
\end{equation}
We suppressed the arguments of the Bell polynomial, and used the Pochhammer symbol $(x)_n = \Gamma(x+n)/\Gamma(x)$. In fact, the Bell polynomials with the 
arguments as above can be evaluated explicitly by use of the exponential generating function,
\begin{equation}
 \exp \left( \sum_{n=1}^\infty \frac{a_n}{n!} x^n \right) = \sum_{n=0}^\infty \frac{B_n(a_1,\dots,a_n)}{n!} x^n \, .
\end{equation}
Inserting the arguments, we get on the left-hand side
\begin{equation}\label{eq:ridiculousequation}
\begin{aligned}
 \exp \left( \sum_{n=1}^\infty \frac{a_n}{n!} x^n \right)  &= \exp \left( \frac{1}{2} \tr \sum_{n=1}^\infty \frac{1}{n} (x T)^n \right) \\
 &= \exp \left( -\frac{1}{2} \tr \ln \left( \mathbbm{1} - x T \right) \right) \\
 &= \left[ \det \left( \mathbbm{1} - x T \right) \right]^{-1/2} \, .
\end{aligned}
\end{equation}
We conclude that the complete Bell polynomials can be obtained by the Taylor expansion coefficients of this expression in $x$ around zero, and thus finally
\begin{equation}\label{eq:ridiculousresult}
 T^{\mu_1\mu_2} \cdots T^{\mu_{2n-1}\mu_{2n}} \eta_{(\mu_1 \mu_2} \cdots \eta_{\mu_{2n-1}\mu_{2n})} = \left. \frac{2^n (d-2)!!}{(d+2(n-1))!!} \partial_x^n 
\frac{1}{\sqrt{\det \left( \mathbbm{1} - x T \right)}} \right|_{x=0} \, .
\end{equation}
Clearly, we can assume that $T$ is traceless - if it is not, we introduce a traceless decomposition in the beginning, then only the traceless part will give 
rise to a nontrivial angular dependence. In four dimensions, we get thus by use of the \CHT{},
\begin{equation}
 \det \left( \mathbbm{1} - x T \right) = 1 - \frac{1}{2} \tau_2 x^2 - \frac{1}{3} \tau_3 x^3 + \left( \det T \right) x^4 \, .
\end{equation}
If the determinant of $T$ is neglected, then the coefficients of the Taylor expansion can be calculated explicitly. A lengthy calculation yields
\begin{equation}
\begin{aligned}
 \frac{1}{\sqrt{1 - \frac{\tau_2}{2} x^2 - \frac{\tau_3}{3} x^3}} &= \sum_{n=0}^\infty \frac{\left( \frac{\tau_2}{2} \right)^{n/2} 
\mathbf{i}^n}{\left( \frac{n-3\mu}{2} \right)!} \pi \left( \frac{\sqrt{2} \tau_3}{3\mathbf{i} \tau_2^{3/2}} \right)^\mu x^n \times \\
&\qquad\quad {}^{}_3 F_2^\text{reg} \left( \frac{2+\mu-n}{6}, \frac{4+\mu-n}{6}, \frac{7\mu-n}{6} ; \mu+\frac{1}{2} , \frac{1+\mu-n}{2} \bigg| 
\frac{6\tau_3^2}{\tau_2^3} \right) \, ,
\end{aligned}
\end{equation}
with $\mu = n \text{ mod } 2$, ${}^{}_3 F_2^\text{reg}$ is the regularised generalised hypergeometric function and $\mathbf{i}^2=-1$.

In fact, one can generalise \eqref{eq:ridiculousresult} to the symmetric contraction of $n$ different tensors:
\begin{equation}\label{eq:mostridiculousresult}
\begin{aligned}
 T_1^{\mu_1\mu_2} \cdots T_n^{\mu_{2n-1}\mu_{2n}} \eta_{(\mu_1 \mu_2} \cdots \eta_{\mu_{2n-1}\mu_{2n})} & \\
=\frac{2^n (d-2)!!}{(d+2(n-1))!!} & \left. \partial_{x_1} \cdots \partial_{x_n} \frac{1}{\sqrt{\det \left( \mathbbm{1} - \sum\limits_{i=1}^n x_i T_i \right)}} \right|_{x_i=0} \, .
\end{aligned}
\end{equation}
This can be proven as follows. Define
\begin{equation}
 \mathcal T = \sum_{i=1}^n x_i T_i \, ,
\end{equation}
then by definition
\begin{equation}
\begin{aligned}
 T_1^{\mu_1\mu_2} \cdots T_n^{\mu_{2n-1}\mu_{2n}} \eta_{(\mu_1 \mu_2} \cdots &\eta_{\mu_{2n-1}\mu_{2n})} \\
 &= \left. \frac{1}{n!} \partial_{x_1} \cdots 
\partial_{x_n} \mathcal T^{\mu_1\mu_2} \cdots \mathcal T^{\mu_{2n-1}\mu_{2n}} \eta_{(\mu_1 \mu_2} \cdots \eta_{\mu_{2n-1}\mu_{2n})} \right|_{x_i=0} \, .
\end{aligned}
\end{equation}
For the right-hand side, we can insert \eqref{eq:ridiculousresult} and obtain
\begin{equation}
\begin{aligned}
 T_1^{\mu_1\mu_2} \cdots T_n^{\mu_{2n-1}\mu_{2n}} \eta_{(\mu_1 \mu_2} \cdots &\eta_{\mu_{2n-1}\mu_{2n})} \\
 &= \left. \frac{1}{n!} \frac{2^n 
(d-2)!!}{(d+2(n-1))!!} \partial_{x_1} \cdots \partial_{x_n} \partial_y^n \frac{1}{\sqrt{\det \left( \mathbbm{1} - y \mathcal T \right)}} \right|_{x_i=y=0} \, .
\end{aligned}
\end{equation}
We can now commute the derivatives freely. Realising that
\begin{equation}
 \frac{1}{\sqrt{\det \left( \mathbbm{1} - y \mathcal T \right)}} = f(y x_1, \dots, y x_n) \, ,
\end{equation}
we find
\begin{equation}
 \left. \partial_{x_1} \cdots \partial_{x_n} \partial_y^n f(y x_1, \dots, y x_n) \right|_{x_i=y=0} = \left. \partial_y^n y^n f^{(1,\dots,1)}(0,\dots,0) 
\right|_{y=0} = n! f^{(1,\dots,1)}(0,\dots,0) \, ,
\end{equation}
and thus follows \eqref{eq:mostridiculousresult}, as claimed.

\bibliographystyle{JHEP}
\bibliography{general_bib}

\providecommand{\href}[2]{#2}\begingroup\raggedright\begin{thebibliography}{100}

\bibitem{Weinberg:1980gg}
S.~Weinberg, \emph{{Ultraviolet divergences in quantum theories of
  gravitation}}, {\emph{General Relativity: An Einstein centenary survey, Eds.
  Hawking, S.W., Israel, W; Cambridge University Press} (1979) 790--831}.

\bibitem{Wetterich:1992yh}
C.~Wetterich, \emph{{Exact evolution equation for the effective potential}},
  \href{https://doi.org/10.1016/0370-2693(93)90726-X}{\emph{Phys.Lett.}
  {\bfseries B301} (1993) 90--94}.

\bibitem{Morris:1993qb}
T.~R. Morris, \emph{{The Exact renormalization group and approximate
  solutions}}, \href{https://doi.org/10.1142/S0217751X94000972}{\emph{Int. J.
  Mod. Phys.} {\bfseries A9} (1994) 2411--2450},
  [\href{https://arxiv.org/abs/hep-ph/9308265}{{\ttfamily hep-ph/9308265}}].

\bibitem{Reuter:1996cp}
M.~Reuter, \emph{{Nonperturbative evolution equation for quantum gravity}},
  \href{https://doi.org/10.1103/PhysRevD.57.971}{\emph{Phys.Rev.} {\bfseries
  D57} (1998) 971--985},
  [\href{https://arxiv.org/abs/hep-th/9605030}{{\ttfamily hep-th/9605030}}].

\bibitem{Falkenberg:1996bq}
S.~Falkenberg and S.~D. Odintsov, \emph{{Gauge dependence of the effective
  average action in Einstein gravity}},
  \href{https://doi.org/10.1142/S0217751X98000263}{\emph{Int. J. Mod. Phys.}
  {\bfseries A13} (1998) 607--623},
  [\href{https://arxiv.org/abs/hep-th/9612019}{{\ttfamily hep-th/9612019}}].

\bibitem{Souma:1999at}
W.~Souma, \emph{{Nontrivial ultraviolet fixed point in quantum gravity}},
  \href{https://doi.org/10.1143/PTP.102.181}{\emph{Prog. Theor. Phys.}
  {\bfseries 102} (1999) 181--195},
  [\href{https://arxiv.org/abs/hep-th/9907027}{{\ttfamily hep-th/9907027}}].

\bibitem{Lauscher:2001ya}
O.~Lauscher and M.~Reuter, \emph{{Ultraviolet fixed point and generalized flow
  equation of quantum gravity}},
  \href{https://doi.org/10.1103/PhysRevD.65.025013}{\emph{Phys.Rev.} {\bfseries
  D65} (2002) 025013}, [\href{https://arxiv.org/abs/hep-th/0108040}{{\ttfamily
  hep-th/0108040}}].

\bibitem{Lauscher:2001rz}
O.~Lauscher and M.~Reuter, \emph{{Is quantum Einstein gravity nonperturbatively
  renormalizable?}},
  \href{https://doi.org/10.1088/0264-9381/19/3/304}{\emph{Class. Quant. Grav.}
  {\bfseries 19} (2002) 483--492},
  [\href{https://arxiv.org/abs/hep-th/0110021}{{\ttfamily hep-th/0110021}}].

\bibitem{Reuter:2001ag}
M.~Reuter and F.~Saueressig, \emph{{Renormalization group flow of quantum
  gravity in the Einstein-Hilbert truncation}},
  \href{https://doi.org/10.1103/PhysRevD.65.065016}{\emph{Phys. Rev.}
  {\bfseries D65} (2002) 065016},
  [\href{https://arxiv.org/abs/hep-th/0110054}{{\ttfamily hep-th/0110054}}].

\bibitem{Litim:2003vp}
D.~F. Litim, \emph{{Fixed points of quantum gravity}},
  \href{https://doi.org/10.1103/PhysRevLett.92.201301}{\emph{Phys. Rev. Lett.}
  {\bfseries 92} (2004) 201301},
  [\href{https://arxiv.org/abs/hep-th/0312114}{{\ttfamily hep-th/0312114}}].

\bibitem{Lauscher:2005qz}
O.~Lauscher and M.~Reuter, \emph{{Fractal spacetime structure in asymptotically
  safe gravity}},
  \href{https://doi.org/10.1088/1126-6708/2005/10/050}{\emph{JHEP} {\bfseries
  10} (2005) 050}, [\href{https://arxiv.org/abs/hep-th/0508202}{{\ttfamily
  hep-th/0508202}}].

\bibitem{Reuter:2005bb}
M.~Reuter and J.-M. Schwindt, \emph{{A Minimal length from the cutoff modes in
  asymptotically safe quantum gravity}},
  \href{https://doi.org/10.1088/1126-6708/2006/01/070}{\emph{JHEP} {\bfseries
  01} (2006) 070}, [\href{https://arxiv.org/abs/hep-th/0511021}{{\ttfamily
  hep-th/0511021}}].

\bibitem{Niedermaier:2006wt}
M.~Niedermaier and M.~Reuter, \emph{{The Asymptotic Safety Scenario in Quantum
  Gravity}}, \href{https://doi.org/10.12942/lrr-2006-5}{\emph{Living Rev.Rel.}
  {\bfseries 9} (2006) 5--173}.

\bibitem{Groh:2010ta}
K.~Groh and F.~Saueressig, \emph{{Ghost wave-function renormalization in
  Asymptotically Safe Quantum Gravity}},
  \href{https://doi.org/10.1088/1751-8113/43/36/365403}{\emph{J. Phys.}
  {\bfseries A43} (2010) 365403},
  [\href{https://arxiv.org/abs/1001.5032}{{\ttfamily 1001.5032}}].

\bibitem{Benedetti:2010nr}
D.~Benedetti, K.~Groh, P.~F. Machado and F.~Saueressig, \emph{{The Universal RG
  Machine}}, \href{https://doi.org/10.1007/JHEP06(2011)079}{\emph{JHEP}
  {\bfseries 1106} (2011) 079},
  [\href{https://arxiv.org/abs/1012.3081}{{\ttfamily 1012.3081}}].

\bibitem{Manrique:2011jc}
E.~Manrique, S.~Rechenberger and F.~Saueressig, \emph{{Asymptotically Safe
  Lorentzian Gravity}},
  \href{https://doi.org/10.1103/PhysRevLett.106.251302}{\emph{Phys. Rev. Lett.}
  {\bfseries 106} (2011) 251302},
  [\href{https://arxiv.org/abs/1102.5012}{{\ttfamily 1102.5012}}].

\bibitem{Reuter:2012id}
M.~Reuter and F.~Saueressig, \emph{{Quantum Einstein Gravity}},
  \href{https://doi.org/10.1088/1367-2630/14/5/055022}{\emph{New J.Phys.}
  {\bfseries 14} (2012) 055022},
  [\href{https://arxiv.org/abs/1202.2274}{{\ttfamily 1202.2274}}].

\bibitem{Harst:2012ni}
U.~Harst and M.~Reuter, \emph{{The 'Tetrad only' theory space: Nonperturbative
  renormalization flow and Asymptotic Safety}},
  \href{https://doi.org/10.1007/JHEP05(2012)005}{\emph{JHEP} {\bfseries 05}
  (2012) 005}, [\href{https://arxiv.org/abs/1203.2158}{{\ttfamily 1203.2158}}].

\bibitem{Litim:2012vz}
D.~Litim and A.~Satz, \emph{{Limit cycles and quantum gravity}},
  \href{https://arxiv.org/abs/1205.4218}{{\ttfamily 1205.4218}}.

\bibitem{Nink:2012kr}
A.~Nink and M.~Reuter, \emph{{On quantum gravity, Asymptotic Safety, and
  paramagnetic dominance}},
  \href{https://doi.org/10.1142/S0218271813300085}{\emph{Int.J.Mod.Phys.}
  {\bfseries D22} (2013) 138--157},
  [\href{https://arxiv.org/abs/1212.4325}{{\ttfamily 1212.4325}}].

\bibitem{Rechenberger:2012dt}
S.~Rechenberger and F.~Saueressig, \emph{{A functional renormalization group
  equation for foliated spacetimes}},
  \href{https://doi.org/10.1007/JHEP03(2013)010}{\emph{JHEP} {\bfseries 03}
  (2013) 010}, [\href{https://arxiv.org/abs/1212.5114}{{\ttfamily 1212.5114}}].

\bibitem{Nink:2014yya}
A.~Nink, \emph{{Field Parametrization Dependence in Asymptotically Safe Quantum
  Gravity}}, \href{https://doi.org/10.1103/PhysRevD.91.044030}{\emph{Phys.Rev.}
  {\bfseries D91} (2015) 044030},
  [\href{https://arxiv.org/abs/1410.7816}{{\ttfamily 1410.7816}}].

\bibitem{Gies:2015tca}
H.~Gies, B.~Knorr and S.~Lippoldt, \emph{{Generalized Parametrization
  Dependence in Quantum Gravity}},
  \href{https://doi.org/10.1103/PhysRevD.92.084020}{\emph{Phys. Rev.}
  {\bfseries D92} (2015) 084020},
  [\href{https://arxiv.org/abs/1507.08859}{{\ttfamily 1507.08859}}].

\bibitem{Falls:2015qga}
K.~Falls, \emph{{On the renormalisation of Newton's constant}},
  \href{https://doi.org/10.1103/PhysRevD.92.124057}{\emph{Phys. Rev.}
  {\bfseries D92} (2015) 124057},
  [\href{https://arxiv.org/abs/1501.05331}{{\ttfamily 1501.05331}}].

\bibitem{Falls:2015cta}
K.~Falls, \emph{{Critical scaling in quantum gravity from the renormalisation
  group}},  \href{https://arxiv.org/abs/1503.06233}{{\ttfamily 1503.06233}}.

\bibitem{Lauscher:2002sq}
O.~Lauscher and M.~Reuter, \emph{{Flow equation of quantum Einstein gravity in
  a higher derivative truncation}},
  \href{https://doi.org/10.1103/PhysRevD.66.025026}{\emph{Phys. Rev.}
  {\bfseries D66} (2002) 025026},
  [\href{https://arxiv.org/abs/hep-th/0205062}{{\ttfamily hep-th/0205062}}].

\bibitem{Codello:2006in}
A.~Codello and R.~Percacci, \emph{{Fixed points of higher derivative gravity}},
  \href{https://doi.org/10.1103/PhysRevLett.97.221301}{\emph{Phys. Rev. Lett.}
  {\bfseries 97} (2006) 221301},
  [\href{https://arxiv.org/abs/hep-th/0607128}{{\ttfamily hep-th/0607128}}].

\bibitem{Benedetti:2009rx}
D.~Benedetti, P.~F. Machado and F.~Saueressig, \emph{{Asymptotic safety in
  higher-derivative gravity}},
  \href{https://doi.org/10.1142/S0217732309031521}{\emph{Mod. Phys. Lett.}
  {\bfseries A24} (2009) 2233--2241},
  [\href{https://arxiv.org/abs/0901.2984}{{\ttfamily 0901.2984}}].

\bibitem{Groh:2011vn}
K.~Groh, S.~Rechenberger, F.~Saueressig and O.~Zanusso, \emph{{Higher
  Derivative Gravity from the Universal Renormalization Group Machine}},
  {\emph{PoS} {\bfseries EPS-HEP2011} (2011) 124},
  [\href{https://arxiv.org/abs/1111.1743}{{\ttfamily 1111.1743}}].

\bibitem{Rechenberger:2012pm}
S.~Rechenberger and F.~Saueressig, \emph{{The $R^2$ phase-diagram of QEG and
  its spectral dimension}},
  \href{https://doi.org/10.1103/PhysRevD.86.024018}{\emph{Phys. Rev.}
  {\bfseries D86} (2012) 024018},
  [\href{https://arxiv.org/abs/1206.0657}{{\ttfamily 1206.0657}}].

\bibitem{Ohta:2013uca}
N.~Ohta and R.~Percacci, \emph{{Higher Derivative Gravity and Asymptotic Safety
  in Diverse Dimensions}},
  \href{https://doi.org/10.1088/0264-9381/31/1/015024}{\emph{Class. Quant.
  Grav.} {\bfseries 31} (2014) 015024},
  [\href{https://arxiv.org/abs/1308.3398}{{\ttfamily 1308.3398}}].

\bibitem{Machado:2007ea}
P.~F. Machado and F.~Saueressig, \emph{{On the renormalization group flow of
  f(R)-gravity}},
  \href{https://doi.org/10.1103/PhysRevD.77.124045}{\emph{Phys.Rev.} {\bfseries
  D77} (2008) 124045}, [\href{https://arxiv.org/abs/0712.0445}{{\ttfamily
  0712.0445}}].

\bibitem{Codello:2007bd}
A.~Codello, R.~Percacci and C.~Rahmede, \emph{{Ultraviolet properties of
  f(R)-gravity}},
  \href{https://doi.org/10.1142/S0217751X08038135}{\emph{Int.J.Mod.Phys.}
  {\bfseries A23} (2008) 143--150},
  [\href{https://arxiv.org/abs/0705.1769}{{\ttfamily 0705.1769}}].

\bibitem{Bonanno:2010bt}
A.~Bonanno, A.~Contillo and R.~Percacci, \emph{{Inflationary solutions in
  asymptotically safe f(R) theories}},
  \href{https://doi.org/10.1088/0264-9381/28/14/145026}{\emph{Class. Quant.
  Grav.} {\bfseries 28} (2011) 145026},
  [\href{https://arxiv.org/abs/1006.0192}{{\ttfamily 1006.0192}}].

\bibitem{Demmel:2012ub}
M.~Demmel, F.~Saueressig and O.~Zanusso, \emph{{Fixed-Functionals of
  three-dimensional Quantum Einstein Gravity}},
  \href{https://doi.org/10.1007/JHEP11(2012)131}{\emph{JHEP} {\bfseries 11}
  (2012) 131}, [\href{https://arxiv.org/abs/1208.2038}{{\ttfamily 1208.2038}}].

\bibitem{Dietz:2012ic}
J.~A. Dietz and T.~R. Morris, \emph{{Asymptotic safety in the f(R)
  approximation}}, \href{https://doi.org/10.1007/JHEP01(2013)108}{\emph{JHEP}
  {\bfseries 1301} (2013) 108},
  [\href{https://arxiv.org/abs/1211.0955}{{\ttfamily 1211.0955}}].

\bibitem{Falls:2013bv}
K.~Falls, D.~Litim, K.~Nikolakopoulos and C.~Rahmede, \emph{{A bootstrap
  towards asymptotic safety}},
  \href{https://arxiv.org/abs/1301.4191}{{\ttfamily 1301.4191}}.

\bibitem{Dietz:2013sba}
J.~A. Dietz and T.~R. Morris, \emph{{Redundant operators in the exact
  renormalisation group and in the f(R) approximation to asymptotic safety}},
  \href{https://doi.org/10.1007/JHEP07(2013)064}{\emph{JHEP} {\bfseries 07}
  (2013) 064}, [\href{https://arxiv.org/abs/1306.1223}{{\ttfamily 1306.1223}}].

\bibitem{Falls:2014tra}
K.~Falls, D.~F. Litim, K.~Nikolakopoulos and C.~Rahmede, \emph{{Further
  evidence for asymptotic safety of quantum gravity}},
  \href{https://doi.org/10.1103/PhysRevD.93.104022}{\emph{Phys. Rev.}
  {\bfseries D93} (2016) 104022},
  [\href{https://arxiv.org/abs/1410.4815}{{\ttfamily 1410.4815}}].

\bibitem{Demmel:2014hla}
M.~Demmel, F.~Saueressig and O.~Zanusso, \emph{{RG flows of Quantum Einstein
  Gravity in the linear-geometric approximation}},
  \href{https://doi.org/10.1016/j.aop.2015.04.018}{\emph{Annals Phys.}
  {\bfseries 359} (2015) 141--165},
  [\href{https://arxiv.org/abs/1412.7207}{{\ttfamily 1412.7207}}].

\bibitem{Demmel:2014sga}
M.~Demmel, F.~Saueressig and O.~Zanusso, \emph{{RG flows of Quantum Einstein
  Gravity on maximally symmetric spaces}},
  \href{https://doi.org/10.1007/JHEP06(2014)026}{\emph{JHEP} {\bfseries 06}
  (2014) 026}, [\href{https://arxiv.org/abs/1401.5495}{{\ttfamily 1401.5495}}].

\bibitem{Eichhorn:2015bna}
A.~Eichhorn, \emph{{The Renormalization Group flow of unimodular f(R)
  gravity}}, \href{https://doi.org/10.1007/JHEP04(2015)096}{\emph{JHEP}
  {\bfseries 1504} (2015) 096},
  [\href{https://arxiv.org/abs/1501.05848}{{\ttfamily 1501.05848}}].

\bibitem{Demmel:2015oqa}
M.~Demmel, F.~Saueressig and O.~Zanusso, \emph{{A proper fixed functional for
  four-dimensional Quantum Einstein Gravity}},
  \href{https://doi.org/10.1007/JHEP08(2015)113}{\emph{JHEP} {\bfseries 08}
  (2015) 113}, [\href{https://arxiv.org/abs/1504.07656}{{\ttfamily
  1504.07656}}].

\bibitem{Ohta:2015efa}
N.~Ohta, R.~Percacci and G.~P. Vacca, \emph{{Flow equation for $f(R)$ gravity
  and some of its exact solutions}},
  \href{https://doi.org/10.1103/PhysRevD.92.061501}{\emph{Phys. Rev.}
  {\bfseries D92} (2015) 061501},
  [\href{https://arxiv.org/abs/1507.00968}{{\ttfamily 1507.00968}}].

\bibitem{Ohta:2015zwa}
N.~Ohta and R.~Percacci, \emph{{Ultraviolet Fixed Points in Conformal Gravity
  and General Quadratic Theories}},
  \href{https://doi.org/10.1088/0264-9381/33/3/035001}{\emph{Class. Quant.
  Grav.} {\bfseries 33} (2016) 035001},
  [\href{https://arxiv.org/abs/1506.05526}{{\ttfamily 1506.05526}}].

\bibitem{Ohta:2015fcu}
N.~Ohta, R.~Percacci and G.~P. Vacca, \emph{{Renormalization Group Equation and
  scaling solutions for f(R) gravity in exponential parametrization}},
  \href{https://doi.org/10.1140/epjc/s10052-016-3895-1}{\emph{Eur. Phys. J.}
  {\bfseries C76} (2016) 46},
  [\href{https://arxiv.org/abs/1511.09393}{{\ttfamily 1511.09393}}].

\bibitem{Falls:2016wsa}
K.~Falls, D.~F. Litim, K.~Nikolakopoulos and C.~Rahmede, \emph{{On de Sitter
  solutions in asymptotically safe $f(R)$ theories}},
  \href{https://arxiv.org/abs/1607.04962}{{\ttfamily 1607.04962}}.

\bibitem{Falls:2016msz}
K.~Falls and N.~Ohta, \emph{{Renormalization Group Equation for $f(R)$ gravity
  on hyperbolic spaces}},
  \href{https://doi.org/10.1103/PhysRevD.94.084005}{\emph{Phys. Rev.}
  {\bfseries D94} (2016) 084005},
  [\href{https://arxiv.org/abs/1607.08460}{{\ttfamily 1607.08460}}].

\bibitem{Morris:2016spn}
T.~R. Morris, \emph{{Large curvature and background scale independence in
  single-metric approximations to asymptotic safety}},
  \href{https://doi.org/10.1007/JHEP11(2016)160}{\emph{JHEP} {\bfseries 11}
  (2016) 160}, [\href{https://arxiv.org/abs/1610.03081}{{\ttfamily
  1610.03081}}].

\bibitem{Christiansen:2016sjn}
N.~Christiansen, \emph{{Four-Derivative Quantum Gravity Beyond Perturbation
  Theory}},  \href{https://arxiv.org/abs/1612.06223}{{\ttfamily 1612.06223}}.

\bibitem{Gonzalez-Martin:2017gza}
S.~Gonzalez-Martin, T.~R. Morris and Z.~H. Slade, \emph{{Asymptotic solutions
  in asymptotic safety}},
  \href{https://doi.org/10.1103/PhysRevD.95.106010}{\emph{Phys. Rev.}
  {\bfseries D95} (2017) 106010},
  [\href{https://arxiv.org/abs/1704.08873}{{\ttfamily 1704.08873}}].

\bibitem{Hamada:2017rvn}
Y.~Hamada and M.~Yamada, \emph{{Asymptotic safety of higher derivative quantum
  gravity non-minimally coupled with a matter system}},
  \href{https://doi.org/10.1007/JHEP08(2017)070}{\emph{JHEP} {\bfseries 08}
  (2017) 070}, [\href{https://arxiv.org/abs/1703.09033}{{\ttfamily
  1703.09033}}].

\bibitem{Nagy:2017zvc}
S.~Nagy, B.~Fazekas, Z.~Peli, K.~Sailer and I.~Steib, \emph{{Regulator
  dependence of fixed points in quantum Einstein gravity with $R^2$
  truncation}},  \href{https://arxiv.org/abs/1707.04934}{{\ttfamily
  1707.04934}}.

\bibitem{Gies:2016con}
H.~Gies, B.~Knorr, S.~Lippoldt and F.~Saueressig, \emph{{Gravitational Two-Loop
  Counterterm Is Asymptotically Safe}},
  \href{https://doi.org/10.1103/PhysRevLett.116.211302}{\emph{Phys. Rev. Lett.}
  {\bfseries 116} (2016) 211302},
  [\href{https://arxiv.org/abs/1601.01800}{{\ttfamily 1601.01800}}].

\bibitem{Nink:2015lmq}
A.~Nink and M.~Reuter, \emph{{The unitary conformal field theory behind 2D
  Asymptotic Safety}},
  \href{https://doi.org/10.1007/JHEP02(2016)167}{\emph{JHEP} {\bfseries 02}
  (2016) 167}, [\href{https://arxiv.org/abs/1512.06805}{{\ttfamily
  1512.06805}}].

\bibitem{Becker:2017tcx}
D.~Becker, C.~Ripken and F.~Saueressig, \emph{{On avoiding Ostrogradski
  instabilities within Asymptotic Safety}},
  \href{https://arxiv.org/abs/1709.09098}{{\ttfamily 1709.09098}}.

\bibitem{Biemans:2016rvp}
J.~Biemans, A.~Platania and F.~Saueressig, \emph{{Quantum gravity on foliated
  spacetimes: Asymptotically safe and sound}},
  \href{https://doi.org/10.1103/PhysRevD.95.086013}{\emph{Phys. Rev.}
  {\bfseries D95} (2017) 086013},
  [\href{https://arxiv.org/abs/1609.04813}{{\ttfamily 1609.04813}}].

\bibitem{Biemans:2017zca}
J.~Biemans, A.~Platania and F.~Saueressig, \emph{{Renormalization group fixed
  points of foliated gravity-matter systems}},
  \href{https://doi.org/10.1007/JHEP05(2017)093}{\emph{JHEP} {\bfseries 05}
  (2017) 093}, [\href{https://arxiv.org/abs/1702.06539}{{\ttfamily
  1702.06539}}].

\bibitem{Houthoff:2017oam}
W.~B. Houthoff, A.~Kurov and F.~Saueressig, \emph{{Impact of topology in
  foliated Quantum Einstein Gravity}},
  \href{https://doi.org/10.1140/epjc/s10052-017-5046-8}{\emph{Eur. Phys. J.}
  {\bfseries C77} (2017) 491},
  [\href{https://arxiv.org/abs/1705.01848}{{\ttfamily 1705.01848}}].

\bibitem{Percacci:2002ie}
R.~Percacci and D.~Perini, \emph{{Constraints on matter from asymptotic
  safety}}, \href{https://doi.org/10.1103/PhysRevD.67.081503}{\emph{Phys. Rev.}
  {\bfseries D67} (2003) 081503},
  [\href{https://arxiv.org/abs/hep-th/0207033}{{\ttfamily hep-th/0207033}}].

\bibitem{Percacci:2003jz}
R.~Percacci and D.~Perini, \emph{{Asymptotic safety of gravity coupled to
  matter}}, \href{https://doi.org/10.1103/PhysRevD.68.044018}{\emph{Phys. Rev.}
  {\bfseries D68} (2003) 044018},
  [\href{https://arxiv.org/abs/hep-th/0304222}{{\ttfamily hep-th/0304222}}].

\bibitem{Zanusso:2009bs}
O.~Zanusso, L.~Zambelli, G.~P. Vacca and R.~Percacci, \emph{{Gravitational
  corrections to Yukawa systems}},
  \href{https://doi.org/10.1016/j.physletb.2010.04.043}{\emph{Phys. Lett.}
  {\bfseries B689} (2010) 90--94},
  [\href{https://arxiv.org/abs/0904.0938}{{\ttfamily 0904.0938}}].

\bibitem{Daum:2009dn}
J.-E. Daum, U.~Harst and M.~Reuter, \emph{{Running Gauge Coupling in
  Asymptotically Safe Quantum Gravity}},
  \href{https://doi.org/10.1007/JHEP01(2010)084}{\emph{JHEP} {\bfseries 01}
  (2010) 084}, [\href{https://arxiv.org/abs/0910.4938}{{\ttfamily 0910.4938}}].

\bibitem{Narain:2009fy}
G.~Narain and R.~Percacci, \emph{{Renormalization Group Flow in Scalar-Tensor
  Theories. I}},
  \href{https://doi.org/10.1088/0264-9381/27/7/075001}{\emph{Class.Quant.Grav.}
  {\bfseries 27} (2010) 075001},
  [\href{https://arxiv.org/abs/0911.0386}{{\ttfamily 0911.0386}}].

\bibitem{Manrique:2010mq}
E.~Manrique, M.~Reuter and F.~Saueressig, \emph{{Matter Induced Bimetric
  Actions for Gravity}},
  \href{https://doi.org/10.1016/j.aop.2010.11.003}{\emph{Annals Phys.}
  {\bfseries 326} (2011) 440--462},
  [\href{https://arxiv.org/abs/1003.5129}{{\ttfamily 1003.5129}}].

\bibitem{Vacca:2010mj}
G.~P. Vacca and O.~Zanusso, \emph{{Asymptotic Safety in Einstein Gravity and
  Scalar-Fermion Matter}},
  \href{https://doi.org/10.1103/PhysRevLett.105.231601}{\emph{Phys. Rev. Lett.}
  {\bfseries 105} (2010) 231601},
  [\href{https://arxiv.org/abs/1009.1735}{{\ttfamily 1009.1735}}].

\bibitem{Harst:2011zx}
U.~Harst and M.~Reuter, \emph{{QED coupled to QEG}},
  \href{https://doi.org/10.1007/JHEP05(2011)119}{\emph{JHEP} {\bfseries 1105}
  (2011) 119}, [\href{https://arxiv.org/abs/1101.6007}{{\ttfamily 1101.6007}}].

\bibitem{Eichhorn:2011pc}
A.~Eichhorn and H.~Gies, \emph{{Light fermions in quantum gravity}},
  \href{https://doi.org/10.1088/1367-2630/13/12/125012}{\emph{New J.Phys.}
  {\bfseries 13} (2011) 125012},
  [\href{https://arxiv.org/abs/1104.5366}{{\ttfamily 1104.5366}}].

\bibitem{Folkerts:2011jz}
S.~Folkerts, D.~F. Litim and J.~M. Pawlowski, \emph{{Asymptotic freedom of
  Yang-Mills theory with gravity}},
  \href{https://doi.org/10.1016/j.physletb.2012.02.002}{\emph{Phys.Lett.}
  {\bfseries B709} (2012) 234--241},
  [\href{https://arxiv.org/abs/1101.5552}{{\ttfamily 1101.5552}}].

\bibitem{Dona:2012am}
P.~Dona and R.~Percacci, \emph{{Functional renormalization with fermions and
  tetrads}}, \href{https://doi.org/10.1103/PhysRevD.87.045002}{\emph{Phys.
  Rev.} {\bfseries D87} (2013) 045002},
  [\href{https://arxiv.org/abs/1209.3649}{{\ttfamily 1209.3649}}].

\bibitem{Dobrich:2012nv}
B.~Dobrich and A.~Eichhorn, \emph{{Can we see quantum gravity? Photons in the
  asymptotic-safety scenario}},
  \href{https://doi.org/10.1007/JHEP06(2012)156}{\emph{JHEP} {\bfseries 06}
  (2012) 156}, [\href{https://arxiv.org/abs/1203.6366}{{\ttfamily 1203.6366}}].

\bibitem{Eichhorn:2012va}
A.~Eichhorn, \emph{{Quantum-gravity-induced matter self-interactions in the
  asymptotic-safety scenario}},
  \href{https://doi.org/10.1103/PhysRevD.86.105021}{\emph{Phys. Rev.}
  {\bfseries D86} (2012) 105021},
  [\href{https://arxiv.org/abs/1204.0965}{{\ttfamily 1204.0965}}].

\bibitem{Dona:2013qba}
P.~Don\`a, A.~Eichhorn and R.~Percacci, \emph{{Matter matters in asymptotically
  safe quantum gravity}},
  \href{https://doi.org/10.1103/PhysRevD.89.084035}{\emph{Phys.Rev.} {\bfseries
  D89} (2014) 084035}, [\href{https://arxiv.org/abs/1311.2898}{{\ttfamily
  1311.2898}}].

\bibitem{Henz:2013oxa}
T.~Henz, J.~M. Pawlowski, A.~Rodigast and C.~Wetterich, \emph{{Dilaton Quantum
  Gravity}},
  \href{https://doi.org/10.1016/j.physletb.2013.10.015}{\emph{Phys.Lett.}
  {\bfseries B727} (2013) 298--302},
  [\href{https://arxiv.org/abs/1304.7743}{{\ttfamily 1304.7743}}].

\bibitem{Eichhorn:2014qka}
A.~Eichhorn and M.~M. Scherer, \emph{{Planck scale, Higgs mass, and scalar dark
  matter}}, \href{https://doi.org/10.1103/PhysRevD.90.025023}{\emph{Phys. Rev.}
  {\bfseries D90} (2014) 025023},
  [\href{https://arxiv.org/abs/1404.5962}{{\ttfamily 1404.5962}}].

\bibitem{Percacci:2015wwa}
R.~Percacci and G.~P. Vacca, \emph{{Search of scaling solutions in
  scalar-tensor gravity}},
  \href{https://doi.org/10.1140/epjc/s10052-015-3410-0}{\emph{Eur. Phys. J.}
  {\bfseries C75} (2015) 188},
  [\href{https://arxiv.org/abs/1501.00888}{{\ttfamily 1501.00888}}].

\bibitem{Borchardt:2015rxa}
J.~Borchardt and B.~Knorr, \emph{{Global solutions of functional fixed point
  equations via pseudospectral methods}},
  \href{https://doi.org/10.1103/PhysRevD.93.089904,
  10.1103/PhysRevD.91.105011}{\emph{Phys. Rev.} {\bfseries D91} (2015) 105011},
  [\href{https://arxiv.org/abs/1502.07511}{{\ttfamily 1502.07511}}].

\bibitem{Dona:2015tnf}
P.~Donà, A.~Eichhorn, P.~Labus and R.~Percacci, \emph{{Asymptotic safety in an
  interacting system of gravity and scalar matter}},
  \href{https://doi.org/10.1103/PhysRevD.93.129904,
  10.1103/PhysRevD.93.044049}{\emph{Phys. Rev.} {\bfseries D93} (2016) 044049},
  [\href{https://arxiv.org/abs/1512.01589}{{\ttfamily 1512.01589}}].

\bibitem{Labus:2015ska}
P.~Labus, R.~Percacci and G.~P. Vacca, \emph{{Asymptotic safety in $O(N)$
  scalar models coupled to gravity}},
  \href{https://doi.org/10.1016/j.physletb.2015.12.022}{\emph{Phys. Lett.}
  {\bfseries B753} (2016) 274--281},
  [\href{https://arxiv.org/abs/1505.05393}{{\ttfamily 1505.05393}}].

\bibitem{Meibohm:2015twa}
J.~Meibohm, J.~M. Pawlowski and M.~Reichert, \emph{{Asymptotic safety of
  gravity-matter systems}},
  \href{https://doi.org/10.1103/PhysRevD.93.084035}{\emph{Phys. Rev.}
  {\bfseries D93} (2016) 084035},
  [\href{https://arxiv.org/abs/1510.07018}{{\ttfamily 1510.07018}}].

\bibitem{Eichhorn:2016esv}
A.~Eichhorn, A.~Held and J.~M. Pawlowski, \emph{{Quantum-gravity effects on a
  Higgs-Yukawa model}},
  \href{https://doi.org/10.1103/PhysRevD.94.104027}{\emph{Phys. Rev.}
  {\bfseries D94} (2016) 104027},
  [\href{https://arxiv.org/abs/1604.02041}{{\ttfamily 1604.02041}}].

\bibitem{Meibohm:2016mkp}
J.~Meibohm and J.~M. Pawlowski, \emph{{Chiral fermions in asymptotically safe
  quantum gravity}},
  \href{https://doi.org/10.1140/epjc/s10052-016-4132-7}{\emph{Eur. Phys. J.}
  {\bfseries C76} (2016) 285},
  [\href{https://arxiv.org/abs/1601.04597}{{\ttfamily 1601.04597}}].

\bibitem{Eichhorn:2016vvy}
A.~Eichhorn and S.~Lippoldt, \emph{{Quantum gravity and Standard-Model-like
  fermions}}, \href{https://doi.org/10.1016/j.physletb.2017.01.064}{\emph{Phys.
  Lett.} {\bfseries B767} (2017) 142--146},
  [\href{https://arxiv.org/abs/1611.05878}{{\ttfamily 1611.05878}}].

\bibitem{Henz:2016aoh}
T.~Henz, J.~M. Pawlowski and C.~Wetterich, \emph{{Scaling solutions for Dilaton
  Quantum Gravity}},
  \href{https://doi.org/10.1016/j.physletb.2017.01.057}{\emph{Phys. Lett.}
  {\bfseries B769} (2017) 105--110},
  [\href{https://arxiv.org/abs/1605.01858}{{\ttfamily 1605.01858}}].

\bibitem{Christiansen:2017gtg}
N.~Christiansen and A.~Eichhorn, \emph{{An asymptotically safe solution to the
  U(1) triviality problem}},
  \href{https://doi.org/10.1016/j.physletb.2017.04.047}{\emph{Phys. Lett.}
  {\bfseries B770} (2017) 154--160},
  [\href{https://arxiv.org/abs/1702.07724}{{\ttfamily 1702.07724}}].

\bibitem{Christiansen:2017qca}
N.~Christiansen, A.~Eichhorn and A.~Held, \emph{{Is scale-invariance in
  gauge-Yukawa systems compatible with the graviton?}},
  \href{https://doi.org/10.1103/PhysRevD.96.084021}{\emph{Phys. Rev.}
  {\bfseries D96} (2017) 084021},
  [\href{https://arxiv.org/abs/1705.01858}{{\ttfamily 1705.01858}}].

\bibitem{Eichhorn:2017eht}
A.~Eichhorn and A.~Held, \emph{{Viability of quantum-gravity induced
  ultraviolet completions for matter}},
  \href{https://arxiv.org/abs/1705.02342}{{\ttfamily 1705.02342}}.

\bibitem{Eichhorn:2017ylw}
A.~Eichhorn and A.~Held, \emph{{Top mass from asymptotic safety}},
  \href{https://arxiv.org/abs/1707.01107}{{\ttfamily 1707.01107}}.

\bibitem{Wetterich:2017ixo}
C.~Wetterich, \emph{{Graviton fluctuations erase the cosmological constant}},
  \href{https://doi.org/10.1016/j.physletb.2017.08.002}{\emph{Phys. Lett.}
  {\bfseries B773} (2017) 6--19},
  [\href{https://arxiv.org/abs/1704.08040}{{\ttfamily 1704.08040}}].

\bibitem{Eichhorn:2017lry}
A.~Eichhorn and F.~Versteegen, \emph{{Upper bound on the Abelian gauge coupling
  from asymptotic safety}},  \href{https://arxiv.org/abs/1709.07252}{{\ttfamily
  1709.07252}}.

\bibitem{Eichhorn:2017sok}
A.~Eichhorn, S.~Lippoldt and V.~Skrinjar, \emph{{Nonminimal hints for
  asymptotic safety}},  \href{https://arxiv.org/abs/1710.03005}{{\ttfamily
  1710.03005}}.

\bibitem{Christiansen:2017cxa}
N.~Christiansen, D.~F. Litim, J.~M. Pawlowski and M.~Reichert, \emph{{One force
  to rule them all: asymptotic safety of gravity with matter}},
  \href{https://arxiv.org/abs/1710.04669}{{\ttfamily 1710.04669}}.

\bibitem{Litim:2002ce}
D.~F. Litim and J.~M. Pawlowski, \emph{{Renormalization group flows for gauge
  theories in axial gauges}},
  \href{https://doi.org/10.1088/1126-6708/2002/09/049}{\emph{JHEP} {\bfseries
  09} (2002) 049}, [\href{https://arxiv.org/abs/hep-th/0203005}{{\ttfamily
  hep-th/0203005}}].

\bibitem{Bridle:2013sra}
I.~H. Bridle, J.~A. Dietz and T.~R. Morris, \emph{{The local potential
  approximation in the background field formalism}},
  \href{https://doi.org/10.1007/JHEP03(2014)093}{\emph{JHEP} {\bfseries 1403}
  (2014) 093}, [\href{https://arxiv.org/abs/1312.2846}{{\ttfamily 1312.2846}}].

\bibitem{Pawlowski:2003sk}
J.~M. Pawlowski, \emph{{Geometrical effective action and Wilsonian flows}},
  \href{https://arxiv.org/abs/hep-th/0310018}{{\ttfamily hep-th/0310018}}.

\bibitem{Pawlowski:2005xe}
J.~M. Pawlowski, \emph{{Aspects of the functional renormalisation group}},
  \href{https://doi.org/10.1016/j.aop.2007.01.007}{\emph{Annals Phys.}
  {\bfseries 322} (2007) 2831--2915},
  [\href{https://arxiv.org/abs/hep-th/0512261}{{\ttfamily hep-th/0512261}}].

\bibitem{Manrique:2009uh}
E.~Manrique and M.~Reuter, \emph{{Bimetric Truncations for Quantum Einstein
  Gravity and Asymptotic Safety}},
  \href{https://doi.org/10.1016/j.aop.2009.11.009}{\emph{Annals Phys.}
  {\bfseries 325} (2010) 785--815},
  [\href{https://arxiv.org/abs/0907.2617}{{\ttfamily 0907.2617}}].

\bibitem{Donkin:2012ud}
I.~Donkin and J.~M. Pawlowski, \emph{{The phase diagram of quantum gravity from
  diffeomorphism-invariant RG-flows}},
  \href{https://arxiv.org/abs/1203.4207}{{\ttfamily 1203.4207}}.

\bibitem{Dietz:2015owa}
J.~A. Dietz and T.~R. Morris, \emph{{Background independent exact
  renormalization group for conformally reduced gravity}},
  \href{https://doi.org/10.1007/JHEP04(2015)118}{\emph{JHEP} {\bfseries 04}
  (2015) 118}, [\href{https://arxiv.org/abs/1502.07396}{{\ttfamily
  1502.07396}}].

\bibitem{Safari:2015dva}
M.~Safari, \emph{{Splitting Ward identity}},
  \href{https://doi.org/10.1140/epjc/s10052-016-4036-6}{\emph{Eur. Phys. J.}
  {\bfseries C76} (2016) 201},
  [\href{https://arxiv.org/abs/1508.06244}{{\ttfamily 1508.06244}}].

\bibitem{Labus:2016lkh}
P.~Labus, T.~R. Morris and Z.~H. Slade, \emph{{Background independence in a
  background dependent renormalization group}},
  \href{https://doi.org/10.1103/PhysRevD.94.024007}{\emph{Phys. Rev.}
  {\bfseries D94} (2016) 024007},
  [\href{https://arxiv.org/abs/1603.04772}{{\ttfamily 1603.04772}}].

\bibitem{Morris:2016nda}
T.~R. Morris and A.~W.~H. Preston, \emph{{Manifestly diffeomorphism invariant
  classical Exact Renormalization Group}},
  \href{https://doi.org/10.1007/JHEP06(2016)012}{\emph{JHEP} {\bfseries 06}
  (2016) 012}, [\href{https://arxiv.org/abs/1602.08993}{{\ttfamily
  1602.08993}}].

\bibitem{Safari:2016dwj}
M.~Safari and G.~P. Vacca, \emph{{Covariant and single-field effective action
  with the background-field formalism}},
  \href{https://doi.org/10.1103/PhysRevD.96.085001}{\emph{Phys. Rev.}
  {\bfseries D96} (2017) 085001},
  [\href{https://arxiv.org/abs/1607.03053}{{\ttfamily 1607.03053}}].

\bibitem{Safari:2016gtj}
M.~Safari and G.~P. Vacca, \emph{{Covariant and background independent
  functional RG flow for the effective average action}},
  \href{https://doi.org/10.1007/JHEP11(2016)139}{\emph{JHEP} {\bfseries 11}
  (2016) 139}, [\href{https://arxiv.org/abs/1607.07074}{{\ttfamily
  1607.07074}}].

\bibitem{Wetterich:2016ewc}
C.~Wetterich, \emph{{Gauge invariant flow equation}},
  \href{https://arxiv.org/abs/1607.02989}{{\ttfamily 1607.02989}}.

\bibitem{Percacci:2016arh}
R.~Percacci and G.~P. Vacca, \emph{{The background scale Ward identity in
  quantum gravity}},
  \href{https://doi.org/10.1140/epjc/s10052-017-4619-x}{\emph{Eur. Phys. J.}
  {\bfseries C77} (2017) 52},
  [\href{https://arxiv.org/abs/1611.07005}{{\ttfamily 1611.07005}}].

\bibitem{Ohta:2017dsq}
N.~Ohta, \emph{{Background Scale Independence in Quantum Gravity}},
  \href{https://doi.org/10.1093/ptep/ptx020}{\emph{PTEP} {\bfseries 2017}
  (2017) 033E02}, [\href{https://arxiv.org/abs/1701.01506}{{\ttfamily
  1701.01506}}].

\bibitem{Nieto:2017ddk}
C.~M. Nieto, R.~Percacci and V.~Skrinjar, \emph{{Split Weyl transformations in
  quantum gravity}},  \href{https://arxiv.org/abs/1708.09760}{{\ttfamily
  1708.09760}}.

\bibitem{Christiansen:2012rx}
N.~Christiansen, D.~F. Litim, J.~M. Pawlowski and A.~Rodigast, \emph{{Fixed
  points and infrared completion of quantum gravity}},
  \href{https://doi.org/10.1016/j.physletb.2013.11.025}{\emph{Phys.Lett.}
  {\bfseries B728} (2014) 114--117},
  [\href{https://arxiv.org/abs/1209.4038}{{\ttfamily 1209.4038}}].

\bibitem{Codello:2013fpa}
A.~Codello, G.~D'Odorico and C.~Pagani, \emph{{Consistent closure of
  renormalization group flow equations in quantum gravity}},
  \href{https://doi.org/10.1103/PhysRevD.89.081701}{\emph{Phys.Rev.} {\bfseries
  D89} (2014) 081701}, [\href{https://arxiv.org/abs/1304.4777}{{\ttfamily
  1304.4777}}].

\bibitem{Christiansen:2014raa}
N.~Christiansen, B.~Knorr, J.~M. Pawlowski and A.~Rodigast, \emph{{Global Flows
  in Quantum Gravity}},
  \href{https://doi.org/10.1103/PhysRevD.93.044036}{\emph{Phys. Rev.}
  {\bfseries D93} (2016) 044036},
  [\href{https://arxiv.org/abs/1403.1232}{{\ttfamily 1403.1232}}].

\bibitem{Christiansen:2015rva}
N.~Christiansen, B.~Knorr, J.~Meibohm, J.~M. Pawlowski and M.~Reichert,
  \emph{{Local Quantum Gravity}},
  \href{https://doi.org/10.1103/PhysRevD.92.121501}{\emph{Phys. Rev.}
  {\bfseries D92} (2015) 121501},
  [\href{https://arxiv.org/abs/1506.07016}{{\ttfamily 1506.07016}}].

\bibitem{Denz:2016qks}
T.~Denz, J.~M. Pawlowski and M.~Reichert, \emph{{Towards apparent convergence
  in asymptotically safe quantum gravity}},
  \href{https://arxiv.org/abs/1612.07315}{{\ttfamily 1612.07315}}.

\bibitem{Knorr:2017fus}
B.~Knorr and S.~Lippoldt, \emph{{Correlation functions on a curved
  background}}, \href{https://doi.org/10.1103/PhysRevD.96.065020}{\emph{Phys.
  Rev.} {\bfseries D96} (2017) 065020},
  [\href{https://arxiv.org/abs/1707.01397}{{\ttfamily 1707.01397}}].

\bibitem{Manrique:2010am}
E.~Manrique, M.~Reuter and F.~Saueressig, \emph{{Bimetric Renormalization Group
  Flows in Quantum Einstein Gravity}},
  \href{https://doi.org/10.1016/j.aop.2010.11.006}{\emph{Annals Phys.}
  {\bfseries 326} (2011) 463--485},
  [\href{https://arxiv.org/abs/1006.0099}{{\ttfamily 1006.0099}}].

\bibitem{Becker:2014qya}
D.~Becker and M.~Reuter, \emph{{En route to Background Independence: Broken
  split-symmetry, and how to restore it with bi-metric average actions}},
  \href{https://doi.org/10.1016/j.aop.2014.07.023}{\emph{Annals Phys.}
  {\bfseries 350} (2014) 225--301},
  [\href{https://arxiv.org/abs/1404.4537}{{\ttfamily 1404.4537}}].

\bibitem{Becker:2014jua}
D.~Becker and M.~Reuter, \emph{{Propagating gravitons vs. 'dark matter` in
  asymptotically safe quantum gravity}},
  \href{https://doi.org/10.1007/JHEP12(2014)025}{\emph{JHEP} {\bfseries 12}
  (2014) 025}, [\href{https://arxiv.org/abs/1407.5848}{{\ttfamily 1407.5848}}].

\bibitem{Wetterich:2016qee}
C.~Wetterich, \emph{{Gauge symmetry from decoupling}},
  \href{https://doi.org/10.1016/j.nuclphysb.2016.12.008}{\emph{Nucl. Phys.}
  {\bfseries B915} (2017) 135--167},
  [\href{https://arxiv.org/abs/1608.01515}{{\ttfamily 1608.01515}}].

\bibitem{Demmel:2015zfa}
M.~Demmel and A.~Nink, \emph{{Connections and geodesics in the space of
  metrics}}, \href{https://doi.org/10.1103/PhysRevD.92.104013}{\emph{Phys.
  Rev.} {\bfseries D92} (2015) 104013},
  [\href{https://arxiv.org/abs/1506.03809}{{\ttfamily 1506.03809}}].

\bibitem{Ohta:2016npm}
N.~Ohta, R.~Percacci and A.~D. Pereira, \emph{{Gauges and functional measures
  in quantum gravity I: Einstein theory}},
  \href{https://doi.org/10.1007/JHEP06(2016)115}{\emph{JHEP} {\bfseries 06}
  (2016) 115}, [\href{https://arxiv.org/abs/1605.00454}{{\ttfamily
  1605.00454}}].

\bibitem{Kawai:1993fq}
H.~Kawai, Y.~Kitazawa and M.~Ninomiya, \emph{{Quantum gravity in
  (2+epsilon)-dimensions}},
  \href{https://doi.org/10.1143/PTPS.114.149}{\emph{Prog. Theor. Phys. Suppl.}
  {\bfseries 114} (1993) 149--174}.

\bibitem{Kawai:1992np}
H.~Kawai, Y.~Kitazawa and M.~Ninomiya, \emph{{Scaling exponents in quantum
  gravity near two-dimensions}},
  \href{https://doi.org/10.1016/0550-3213(93)90246-L}{\emph{Nucl. Phys.}
  {\bfseries B393} (1993) 280--300},
  [\href{https://arxiv.org/abs/hep-th/9206081}{{\ttfamily hep-th/9206081}}].

\bibitem{Kawai:1993mb}
H.~Kawai, Y.~Kitazawa and M.~Ninomiya, \emph{{Ultraviolet stable fixed point
  and scaling relations in (2+epsilon)-dimensional quantum gravity}},
  \href{https://doi.org/10.1016/0550-3213(93)90594-F}{\emph{Nucl. Phys.}
  {\bfseries B404} (1993) 684--716},
  [\href{https://arxiv.org/abs/hep-th/9303123}{{\ttfamily hep-th/9303123}}].

\bibitem{Kawai:1995ju}
H.~Kawai, Y.~Kitazawa and M.~Ninomiya, \emph{{Renormalizability of quantum
  gravity near two-dimensions}},
  \href{https://doi.org/10.1016/0550-3213(96)00119-8}{\emph{Nucl. Phys.}
  {\bfseries B467} (1996) 313--331},
  [\href{https://arxiv.org/abs/hep-th/9511217}{{\ttfamily hep-th/9511217}}].

\bibitem{Aida:1994zc}
T.~Aida, Y.~Kitazawa, H.~Kawai and M.~Ninomiya, \emph{{Conformal invariance and
  renormalization group in quantum gravity near two-dimensions}},
  \href{https://doi.org/10.1016/0550-3213(94)90273-9}{\emph{Nucl. Phys.}
  {\bfseries B427} (1994) 158--180},
  [\href{https://arxiv.org/abs/hep-th/9404171}{{\ttfamily hep-th/9404171}}].

\bibitem{Manrique:2008zw}
E.~Manrique and M.~Reuter, \emph{{Bare Action and Regularized Functional
  Integral of Asymptotically Safe Quantum Gravity}},
  \href{https://doi.org/10.1103/PhysRevD.79.025008}{\emph{Phys. Rev.}
  {\bfseries D79} (2009) 025008},
  [\href{https://arxiv.org/abs/0811.3888}{{\ttfamily 0811.3888}}].

\bibitem{Manrique:2009tj}
E.~Manrique and M.~Reuter, \emph{{Bare versus Effective Fixed Point Action in
  Asymptotic Safety: The Reconstruction Problem}}, {\emph{PoS} {\bfseries
  CLAQG08} (2011) 001}, [\href{https://arxiv.org/abs/0905.4220}{{\ttfamily
  0905.4220}}].

\bibitem{Morris:2015oca}
T.~R. Morris and Z.~H. Slade, \emph{{Solutions to the reconstruction problem in
  asymptotic safety}},
  \href{https://doi.org/10.1007/JHEP11(2015)094}{\emph{JHEP} {\bfseries 11}
  (2015) 094}, [\href{https://arxiv.org/abs/1507.08657}{{\ttfamily
  1507.08657}}].

\bibitem{Reuter:1996ub}
M.~Reuter, \emph{{Effective average actions and nonperturbative evolution
  equations}},  \href{https://arxiv.org/abs/hep-th/9602012}{{\ttfamily
  hep-th/9602012}}.

\bibitem{Percacci:2007sz}
R.~Percacci, \emph{{Asymptotic Safety}},
  \href{https://arxiv.org/abs/0709.3851}{{\ttfamily 0709.3851}}.

\bibitem{Nagy:2012ef}
S.~Nagy, \emph{{Lectures on renormalization and asymptotic safety}},
  \href{https://doi.org/10.1016/j.aop.2014.07.027}{\emph{Annals Phys.}
  {\bfseries 350} (2014) 310--346},
  [\href{https://arxiv.org/abs/1211.4151}{{\ttfamily 1211.4151}}].

\bibitem{Reuter:2008wj}
M.~Reuter and H.~Weyer, \emph{{Background Independence and Asymptotic Safety in
  Conformally Reduced Gravity}},
  \href{https://doi.org/10.1103/PhysRevD.79.105005}{\emph{Phys. Rev.}
  {\bfseries D79} (2009) 105005},
  [\href{https://arxiv.org/abs/0801.3287}{{\ttfamily 0801.3287}}].

\bibitem{Reuter:2008qx}
M.~Reuter and H.~Weyer, \emph{{Conformal sector of Quantum Einstein Gravity in
  the local potential approximation: Non-Gaussian fixed point and a phase of
  unbroken diffeomorphism invariance}},
  \href{https://doi.org/10.1103/PhysRevD.80.025001}{\emph{Phys. Rev.}
  {\bfseries D80} (2009) 025001},
  [\href{https://arxiv.org/abs/0804.1475}{{\ttfamily 0804.1475}}].

\bibitem{Machado:2009ph}
P.~F. Machado and R.~Percacci, \emph{{Conformally reduced quantum gravity
  revisited}}, \href{https://doi.org/10.1103/PhysRevD.80.024020}{\emph{Phys.
  Rev.} {\bfseries D80} (2009) 024020},
  [\href{https://arxiv.org/abs/0904.2510}{{\ttfamily 0904.2510}}].

\bibitem{Bonanno:2012dg}
A.~Bonanno and F.~Guarnieri, \emph{{Universality and Symmetry Breaking in
  Conformally Reduced Quantum Gravity}},
  \href{https://doi.org/10.1103/PhysRevD.86.105027}{\emph{Phys. Rev.}
  {\bfseries D86} (2012) 105027},
  [\href{https://arxiv.org/abs/1206.6531}{{\ttfamily 1206.6531}}].

\bibitem{Dietz:2016gzg}
J.~A. Dietz, T.~R. Morris and Z.~H. Slade, \emph{{Fixed point structure of the
  conformal factor field in quantum gravity}},
  \href{https://doi.org/10.1103/PhysRevD.94.124014}{\emph{Phys. Rev.}
  {\bfseries D94} (2016) 124014},
  [\href{https://arxiv.org/abs/1605.07636}{{\ttfamily 1605.07636}}].

\bibitem{Litim:1998qi}
D.~F. Litim and J.~M. Pawlowski, \emph{{Flow equations for Yang-Mills theories
  in general axial gauges}},
  \href{https://doi.org/10.1016/S0370-2693(98)00761-8}{\emph{Phys.Lett.}
  {\bfseries B435} (1998) 181--188},
  [\href{https://arxiv.org/abs/hep-th/9802064}{{\ttfamily hep-th/9802064}}].

\bibitem{xActwebpage}
``{xAct: Efficient tensor computer algebra for Mathematica}.''
  \url{http://xact.es/index.html}.

\bibitem{Brizuela:2008ra}
D.~Brizuela, J.~M. Martin-Garcia and G.~A. Mena~Marugan, \emph{{xPert: Computer
  algebra for metric perturbation theory}},
  \href{https://doi.org/10.1007/s10714-009-0773-2}{\emph{Gen. Rel. Grav.}
  {\bfseries 41} (2009) 2415--2431},
  [\href{https://arxiv.org/abs/0807.0824}{{\ttfamily 0807.0824}}].

\bibitem{2008CoPhC.179..597M}
J.~M. {Mart{\'{\i}}n-Garc{\'{\i}}a}, \emph{{xPerm: fast index canonicalization
  for tensor computer algebra}},
  \href{https://doi.org/10.1016/j.cpc.2008.05.009}{\emph{Computer Physics
  Communications} {\bfseries 179} (Oct., 2008) 597--603},
  [\href{https://arxiv.org/abs/0803.0862}{{\ttfamily 0803.0862}}].

\bibitem{2007CoPhC.177..640M}
J.~M. {Mart{\'{\i}}n-Garc{\'{\i}}a}, R.~{Portugal} and L.~R.~U. {Manssur},
  \emph{{The Invar tensor package}},
  \href{https://doi.org/10.1016/j.cpc.2007.05.015}{\emph{Computer Physics
  Communications} {\bfseries 177} (Oct., 2007) 640--648},
  [\href{https://arxiv.org/abs/0704.1756}{{\ttfamily 0704.1756}}].

\bibitem{2008CoPhC.179..586M}
J.~M. {Mart{\'{\i}}n-Garc{\'{\i}}a}, D.~{Yllanes} and R.~{Portugal}, \emph{{The
  Invar tensor package: Differential invariants of Riemann}},
  \href{https://doi.org/10.1016/j.cpc.2008.04.018}{\emph{Computer Physics
  Communications} {\bfseries 179} (Oct., 2008) 586--590},
  [\href{https://arxiv.org/abs/0802.1274}{{\ttfamily 0802.1274}}].

\bibitem{2014CoPhC.185.1719N}
T.~{Nutma}, \emph{{xTras: A field-theory inspired xAct package for
  mathematica}},
  \href{https://doi.org/10.1016/j.cpc.2014.02.006}{\emph{Computer Physics
  Communications} {\bfseries 185} (June, 2014) 1719--1738},
  [\href{https://arxiv.org/abs/1308.3493}{{\ttfamily 1308.3493}}].

\bibitem{Litim:2001up}
D.~F. Litim, \emph{{Optimized renormalization group flows}},
  \href{https://doi.org/10.1103/PhysRevD.64.105007}{\emph{Phys.Rev.} {\bfseries
  D64} (2001) 105007}, [\href{https://arxiv.org/abs/hep-th/0103195}{{\ttfamily
  hep-th/0103195}}].

\bibitem{Litim:2002cf}
D.~F. Litim, \emph{{Critical exponents from optimized renormalization group
  flows}},
  \href{https://doi.org/10.1016/S0550-3213(02)00186-4}{\emph{Nucl.Phys.}
  {\bfseries B631} (2002) 128--158},
  [\href{https://arxiv.org/abs/hep-th/0203006}{{\ttfamily hep-th/0203006}}].

\bibitem{Borchardt:2016pif}
J.~Borchardt and B.~Knorr, \emph{{Solving functional flow equations with
  pseudo-spectral methods}},
  \href{https://doi.org/10.1103/PhysRevD.94.025027}{\emph{Phys. Rev.}
  {\bfseries D94} (2016) 025027},
  [\href{https://arxiv.org/abs/1603.06726}{{\ttfamily 1603.06726}}].

\bibitem{Heilmann:2014iga}
M.~Heilmann, T.~Hellwig, B.~Knorr, M.~Ansorg and A.~Wipf, \emph{{Convergence of
  Derivative Expansion in Supersymmetric Functional RG Flows}},
  \href{https://doi.org/10.1007/JHEP02(2015)109}{\emph{JHEP} {\bfseries 1502}
  (2015) 109}, [\href{https://arxiv.org/abs/1409.5650}{{\ttfamily 1409.5650}}].

\bibitem{Borchardt:2016xju}
J.~Borchardt, H.~Gies and R.~Sondenheimer, \emph{{Global flow of the Higgs
  potential in a Yukawa model}},
  \href{https://doi.org/10.1140/epjc/s10052-016-4300-9}{\emph{Eur. Phys. J.}
  {\bfseries C76} (2016) 472},
  [\href{https://arxiv.org/abs/1603.05861}{{\ttfamily 1603.05861}}].

\bibitem{Borchardt:2016kco}
J.~Borchardt and A.~Eichhorn, \emph{{Universal behavior of coupled order
  parameters below three dimensions}},
  \href{https://doi.org/10.1103/PhysRevE.94.042105}{\emph{Phys. Rev.}
  {\bfseries E94} (2016) 042105},
  [\href{https://arxiv.org/abs/1606.07449}{{\ttfamily 1606.07449}}].

\bibitem{Knorr:2016sfs}
B.~Knorr, \emph{{Ising and Gross-Neveu model in next-to-leading order}},
  \href{https://doi.org/10.1103/PhysRevB.94.245102}{\emph{Phys. Rev.}
  {\bfseries B94} (2016) 245102},
  [\href{https://arxiv.org/abs/1609.03824}{{\ttfamily 1609.03824}}].

\bibitem{Knorr:2017yze}
B.~Knorr, \emph{{Critical (Chiral) Heisenberg Model with the Functional
  Renormalisation Group}},  \href{https://arxiv.org/abs/1708.06200}{{\ttfamily
  1708.06200}}.

\bibitem{Ambjorn:1998xu}
J.~Ambjorn and R.~Loll, \emph{{Nonperturbative Lorentzian quantum gravity,
  causality and topology change}},
  \href{https://doi.org/10.1016/S0550-3213(98)00692-0}{\emph{Nucl. Phys.}
  {\bfseries B536} (1998) 407--434},
  [\href{https://arxiv.org/abs/hep-th/9805108}{{\ttfamily hep-th/9805108}}].

\bibitem{Ambjorn:2005db}
J.~Ambjorn, J.~Jurkiewicz and R.~Loll, \emph{{Spectral dimension of the
  universe}}, \href{https://doi.org/10.1103/PhysRevLett.95.171301}{\emph{Phys.
  Rev. Lett.} {\bfseries 95} (2005) 171301},
  [\href{https://arxiv.org/abs/hep-th/0505113}{{\ttfamily hep-th/0505113}}].

\bibitem{Ambjorn:2006jf}
J.~Ambjorn, J.~Jurkiewicz and R.~Loll, \emph{{Quantum Gravity, or The Art of
  Building Spacetime}},  \href{https://arxiv.org/abs/hep-th/0604212}{{\ttfamily
  hep-th/0604212}}.

\bibitem{Benedetti:2009ge}
D.~Benedetti and J.~Henson, \emph{{Spectral geometry as a probe of quantum
  spacetime}}, \href{https://doi.org/10.1103/PhysRevD.80.124036}{\emph{Phys.
  Rev.} {\bfseries D80} (2009) 124036},
  [\href{https://arxiv.org/abs/0911.0401}{{\ttfamily 0911.0401}}].

\bibitem{Anderson:2011bj}
C.~Anderson, S.~J. Carlip, J.~H. Cooperman, P.~Horava, R.~K. Kommu and P.~R.
  Zulkowski, \emph{{Quantizing Horava-Lifshitz Gravity via Causal Dynamical
  Triangulations}}, \href{https://doi.org/10.1103/PhysRevD.85.044027,
  10.1103/PhysRevD.85.049904}{\emph{Phys. Rev.} {\bfseries D85} (2012) 044027},
  [\href{https://arxiv.org/abs/1111.6634}{{\ttfamily 1111.6634}}].

\bibitem{Ambjorn:2011cg}
J.~Ambjorn, S.~Jordan, J.~Jurkiewicz and R.~Loll, \emph{{A Second-order phase
  transition in CDT}},
  \href{https://doi.org/10.1103/PhysRevLett.107.211303}{\emph{Phys. Rev. Lett.}
  {\bfseries 107} (2011) 211303},
  [\href{https://arxiv.org/abs/1108.3932}{{\ttfamily 1108.3932}}].

\bibitem{Laiho:2011ya}
J.~Laiho and D.~Coumbe, \emph{{Evidence for Asymptotic Safety from Lattice
  Quantum Gravity}},
  \href{https://doi.org/10.1103/PhysRevLett.107.161301}{\emph{Phys. Rev. Lett.}
  {\bfseries 107} (2011) 161301},
  [\href{https://arxiv.org/abs/1104.5505}{{\ttfamily 1104.5505}}].

\bibitem{Ambjorn:2012jv}
J.~Ambjorn, A.~Goerlich, J.~Jurkiewicz and R.~Loll, \emph{{Nonperturbative
  Quantum Gravity}},
  \href{https://doi.org/10.1016/j.physrep.2012.03.007}{\emph{Phys. Rept.}
  {\bfseries 519} (2012) 127--210},
  [\href{https://arxiv.org/abs/1203.3591}{{\ttfamily 1203.3591}}].

\bibitem{Jordan:2013awa}
S.~Jordan and R.~Loll, \emph{{Causal Dynamical Triangulations without Preferred
  Foliation}},
  \href{https://doi.org/10.1016/j.physletb.2013.06.007}{\emph{Phys. Lett.}
  {\bfseries B724} (2013) 155--159},
  [\href{https://arxiv.org/abs/1305.4582}{{\ttfamily 1305.4582}}].

\bibitem{Ambjorn:2013tki}
J.~Ambjørn, A.~Görlich, J.~Jurkiewicz and R.~Loll, \emph{{Quantum Gravity via
  Causal Dynamical Triangulations}},  in \emph{Springer Handbook of Spacetime}
  (A.~Ashtekar and V.~Petkov, eds.), pp.~723--741.
\newblock 2014.
\newblock \href{https://arxiv.org/abs/1302.2173}{{\ttfamily 1302.2173}}.
\newblock \href{https://doi.org/10.1007/978-3-642-41992-8_34}{DOI}.

\bibitem{Ambjorn:2013apa}
J.~Ambjorn, A.~Görlich, J.~Jurkiewicz and R.~Loll, \emph{{Causal dynamical
  triangulations and the search for a theory of quantum gravity}},
  \href{https://doi.org/10.1142/S021827181330019X}{\emph{Int. J. Mod. Phys.}
  {\bfseries D22} (2013) 1330019}.

\bibitem{Ambjorn:2014gsa}
J.~Ambjorn, A.~Görlich, J.~Jurkiewicz, A.~Kreienbuehl and R.~Loll,
  \emph{{Renormalization Group Flow in CDT}},
  \href{https://doi.org/10.1088/0264-9381/31/16/165003}{\emph{Class. Quant.
  Grav.} {\bfseries 31} (2014) 165003},
  [\href{https://arxiv.org/abs/1405.4585}{{\ttfamily 1405.4585}}].

\bibitem{Cooperman:2014uha}
J.~H. Cooperman, \emph{{Making the case for causal dynamical triangulations}},
  \href{https://arxiv.org/abs/1410.0670}{{\ttfamily 1410.0670}}.

\bibitem{Ambjorn:2016mnn}
J.~Ambjørn, J.~Gizbert-Studnicki, A.~Görlich, J.~Jurkiewicz, N.~Klitgaard and
  R.~Loll, \emph{{Characteristics of the new phase in CDT}},
  \href{https://doi.org/10.1140/epjc/s10052-017-4710-3}{\emph{Eur. Phys. J.}
  {\bfseries C77} (2017) 152},
  [\href{https://arxiv.org/abs/1610.05245}{{\ttfamily 1610.05245}}].

\bibitem{Glaser:2016smx}
L.~Glaser, T.~P. Sotiriou and S.~Weinfurtner, \emph{{Extrinsic curvature in
  two-dimensional causal dynamical triangulation}},
  \href{https://doi.org/10.1103/PhysRevD.94.064014}{\emph{Phys. Rev.}
  {\bfseries D94} (2016) 064014},
  [\href{https://arxiv.org/abs/1605.09618}{{\ttfamily 1605.09618}}].

\bibitem{Laiho:2016nlp}
J.~Laiho, S.~Bassler, D.~Coumbe, D.~Du and J.~T. Neelakanta, \emph{{Lattice
  Quantum Gravity and Asymptotic Safety}},
  \href{https://doi.org/10.1103/PhysRevD.96.064015}{\emph{Phys. Rev.}
  {\bfseries D96} (2017) 064015},
  [\href{https://arxiv.org/abs/1604.02745}{{\ttfamily 1604.02745}}].

\bibitem{Ambjorn:2017tnl}
J.~Ambjorn, D.~Coumbe, J.~Gizbert-Studnicki, A.~Gorlich and J.~Jurkiewicz,
  \emph{{New higher-order transition in causal dynamical triangulations}},
  \href{https://doi.org/10.1103/PhysRevD.95.124029}{\emph{Phys. Rev.}
  {\bfseries D95} (2017) 124029},
  [\href{https://arxiv.org/abs/1704.04373}{{\ttfamily 1704.04373}}].

\bibitem{Ambjorn:2017ogo}
J.~Ambjørn, J.~Gizbert-Studnicki, A.~Görlich, K.~Grosvenor and J.~Jurkiewicz,
  \emph{{Four-dimensional CDT with toroidal topology}},
  \href{https://doi.org/10.1016/j.nuclphysb.2017.06.026}{\emph{Nucl. Phys.}
  {\bfseries B922} (2017) 226--246},
  [\href{https://arxiv.org/abs/1705.07653}{{\ttfamily 1705.07653}}].

\bibitem{Glaser:2017yxz}
L.~Glaser and R.~Loll, \emph{{CDT and Cosmology}},
  \href{https://doi.org/10.1016/j.crhy.2017.04.002}{\emph{Comptes Rendus
  Physique} {\bfseries 18} (2017) 265--274},
  [\href{https://arxiv.org/abs/1703.08160}{{\ttfamily 1703.08160}}].

\bibitem{Bombelli:1987aa}
L.~Bombelli, J.~Lee, D.~Meyer and R.~Sorkin, \emph{{Space-Time as a Causal
  Set}}, \href{https://doi.org/10.1103/PhysRevLett.59.521}{\emph{Phys. Rev.
  Lett.} {\bfseries 59} (1987) 521--524}.

\bibitem{Sorkin:1990bj}
R.~D. Sorkin, \emph{{Space-time and causal sets}},  in \emph{{Proceedings, 7th
  Latin American International Symposium on General Relativity and Gravitation:
  Cocoyoc, Mexico, December 2-7, 1990}}, 1990.

\bibitem{Sorkin:2003bx}
R.~D. Sorkin, \emph{{Causal sets: Discrete gravity}},  in \emph{{Lectures on
  quantum gravity. Proceedings, School of Quantum Gravity, Valdivia, Chile,
  January 4-14, 2002}}, pp.~305--327, 2003,
  \href{https://arxiv.org/abs/gr-qc/0309009}{{\ttfamily gr-qc/0309009}},
  \href{https://doi.org/10.1007/0-387-24992-3_7}{DOI}.

\bibitem{Surya:2011du}
S.~Surya, \emph{{Evidence for a Phase Transition in 2D Causal Set Quantum
  Gravity}}, \href{https://doi.org/10.1088/0264-9381/29/13/132001}{\emph{Class.
  Quant. Grav.} {\bfseries 29} (2012) 132001},
  [\href{https://arxiv.org/abs/1110.6244}{{\ttfamily 1110.6244}}].

\bibitem{Surya:2011yh}
S.~Surya, \emph{{Directions in Causal Set Quantum Gravity}},
  \href{https://arxiv.org/abs/1103.6272}{{\ttfamily 1103.6272}}.

\bibitem{Glaser:2013pca}
L.~Glaser and S.~Surya, \emph{{Towards a Definition of Locality in a
  Manifoldlike Causal Set}},
  \href{https://doi.org/10.1103/PhysRevD.88.124026}{\emph{Phys. Rev.}
  {\bfseries D88} (2013) 124026},
  [\href{https://arxiv.org/abs/1309.3403}{{\ttfamily 1309.3403}}].

\bibitem{Glaser:2014dwa}
L.~Glaser and S.~Surya, \emph{{The Hartle–Hawking wave function in 2D causal
  set quantum gravity}},
  \href{https://doi.org/10.1088/0264-9381/33/6/065003}{\emph{Class. Quant.
  Grav.} {\bfseries 33} (2016) 065003},
  [\href{https://arxiv.org/abs/1410.8775}{{\ttfamily 1410.8775}}].

\bibitem{Eichhorn:2017djq}
A.~Eichhorn, S.~Mizera and S.~Surya, \emph{{Echoes of Asymptotic Silence in
  Causal Set Quantum Gravity}},
  \href{https://doi.org/10.1088/1361-6382/aa7d1b}{\emph{Class. Quant. Grav.}
  {\bfseries 34} (2017) 16LT01},
  [\href{https://arxiv.org/abs/1703.08454}{{\ttfamily 1703.08454}}].

\bibitem{Glaser:2017sbe}
L.~Glaser, D.~O'Connor and S.~Surya, \emph{{Finite Size Scaling in 2d Causal
  Set Quantum Gravity}},  \href{https://arxiv.org/abs/1706.06432}{{\ttfamily
  1706.06432}}.

\bibitem{Eichhorn:2017bwe}
A.~Eichhorn, \emph{{Towards coarse graining of discrete Lorentzian quantum
  gravity}},  \href{https://arxiv.org/abs/1709.10419}{{\ttfamily 1709.10419}}.

\bibitem{Hamber:2017pli}
H.~W. Hamber, \emph{{Vacuum Condensate Picture of Quantum Gravity}},
  \href{https://arxiv.org/abs/1707.08188}{{\ttfamily 1707.08188}}.

\bibitem{dbt_mods_00032999}
B.~Knorr, \emph{Asymptotic safety in QFT: from quantum gravity to graphene},
  Ph.D. thesis, Jena, Oct, 2017.

\end{thebibliography}\endgroup

\end{document}